\documentclass[aps,pre,twocolumn,groupedaddress,floatfix]{revtex4-1}

\usepackage{graphicx}
\usepackage[dvipsnames]{xcolor}
\usepackage{amsmath,amssymb}
\usepackage{mathtools}
\usepackage{natbib}
\usepackage[pdftex]{hyperref}
\usepackage{hyperref}
\hypersetup{
    colorlinks=true,
    citecolor=blue,
    linkcolor=blue,
    urlcolor=blue,
}

\allowdisplaybreaks

\begin{document}

\title{Dynamics of ribosomes in mRNA translation under steady and non-steady state conditions}

\author{Juraj Szavits-Nossan}
\email{jszavits@staffmail.ed.ac.uk}
\affiliation{SUPA, School of Physics and Astronomy, University of Edinburgh, Peter Guthrie Tait Road, Edinburgh EH9 3FD, United Kingdom}

\author{Martin R. Evans}
\email{m.evans@ed.ac.uk}
\affiliation{SUPA, School of Physics and Astronomy, University of Edinburgh, Peter Guthrie Tait Road, Edinburgh EH9 3FD, United Kingdom}

\date{\today}

\begin{abstract}
Recent advances in DNA sequencing and fluorescence imaging have made it possible to monitor the dynamics of ribosomes actively engaged in messenger RNA (mRNA) translation.  Here, we model these experiments within the inhomogeneous totally asymmetric simple exclusion process (TASEP) using realistic kinetic parameters. In particular we present analytic expressions to describe the following three cases: (a) translation of a newly transcribed mRNA, (b) translation in the steady state and, specifically the dynamics of individual (tagged) ribosomes and (c) run-off translation after inhibition of translation initiation.  In the cases (b) and (c) we develop an effective medium approximation to describe many-ribosome dynamics in terms of a single tagged ribosome in an effective medium. The predictions are in good agreement with stochastic simulations.
\end{abstract}


\maketitle

\section{\label{Introduction} Introduction}

Protein synthesis is an essential process in all living cells. Proteins are produced by ribosomes from mRNA molecules in a process called translation. A major goal in molecular biology is to understand how the dynamics of translation is influenced by the underlying mRNA sequence. Translation is a complex process that proceeds in three phases: initiation, elongation and termination. A ribosome assembles on the mRNA and initiates translation by recognising the start codon (initiation). After initiation, the ribosome moves along the mRNA molecule in a 5' to 3' direction and assembles the amino acid chain by adding one amino acid for each codon on the mRNA sequence (elongation), until it recognises the stop codon and releases the final protein (termination).

Ribosome movement along the mRNA  has been shown experimentally to be non-uniform \cite{Varenne1984} and this has been linked to the availability of the transfer RNA (tRNA) molecules delivering the correct amino acid to the ribosome \cite{Ikemura1985}. The differences between populations of isoaccepting tRNAs (tRNAs that deliver the same amino acid) correlate with codon usage bias, a phenomenon of non-uniform usage of synonymous codons that code for the same amino acid \cite{Sharp1987}. The idea that the same protein can be translated more efficiently depending on the choice of synonymous codons has been used to increase the production of proteins that are non-native to their host cell \cite{Gustafsson2004}. Despite these successes, others have demonstrated that translation is mostly rate-limited by initiation and codon composition has a lesser effect on protein production under normal cellular conditions \cite{Kudla2009,Shah2013,Cambray2018}.  Thus the issue of how the rate of translation, and hence protein production, is fine-tuned by the underlying genetic sequence remains hotly debated.

A simple theoretical model, known as the totally asymmetric simple exclusion process (TASEP), has been used extensively to understand the dynamics of translation \cite{MacDonald1968,MacDonald1969}. The TASEP captures stochastic motion of individual ribosomes on the mRNA and accounts for excluded-volume interactions between ribosomes that may lead to traffic jams. There is a large body of work on the TASEP applied to mRNA translation and many biological details have been added to improve the original model \cite{vonderHaar2012,Zur2016}. Outside of the biological context, the TASEP has been widely studied in mathematics in the theory of interacting particle systems \cite{Spitzer1970} where it got its name, and in physics as one of the simplest models of transport far from the thermal equilibrium and nonequilibrium statistical physics generally \cite{SchadSchneider2010,Krapivsky2010,Chou2011}. Usually the homogeneous case, which corresponds to uniform elongation rate for ribosomes, is considered and many exact results have been obtained \cite{Derrida1993,Schuetz1993,Schuetz1997,Sasamoto1998,Blythe2007}. The inhomogeneous case, which  corresponds to codon-specific elongation, remains a challenging problem and one must resort to simulations and approximations to make predictions \cite{Shaw2003,Chou2004,Szavits2013,Szavits2018,Szavits2018b,ErdmannPham2020}. 

On the experimental side, in recent years several new techniques have been developed to directly monitor translation kinetics. Ribosome profiling (or Ribo-seq) is a technique based on DNA sequencing of ribosome-protected mRNA fragments that captures the positions of all ribosomes bound to the mRNA at a given time \cite{Ingolia2009}. Translation kinetics is monitored after treating cells with harringtonine, a drug that inhibits new translation initiation. Ribosome profiling experiments are repeated at different times and the average elongation rate is inferred from the linear decrease in the number of ribosome-protected fragments over time \cite{Ingolia2011,Dana2012}. A disadvantage of this method is that it requires averaging over many cells that must be lysed before the measurement is taken, meaning that the information about ribosome dynamics on individual mRNAs is lost.

A direct method of probing dynamics of translation on individual mRNAs in real time is fluorescence imaging of ribosomes tagged with green fluorescent proteins (GFPs) \cite{Yan2016}. The tagging system is achieved by inserting a sequence of $24$ SunTag peptides upstream of the gene of interest. Once translated by a ribosome, these peptides have a high affinity for GFPs resulting in a enhanced fluorescence signal at the ribosome's position. At a newly transcribed mRNA, the fluorescence signal increases linearly over time until the steady state is reached. After treating cells with harringtonine that stops new initiation, the remaining ribosomes run off the mRNA and the average elongation rate is estimated from the linear decay of the fluorescence signal---we will refer to this regime as run-off translation.

\begin{figure}[hbt]
	\centering
	\includegraphics[width=3.4in]{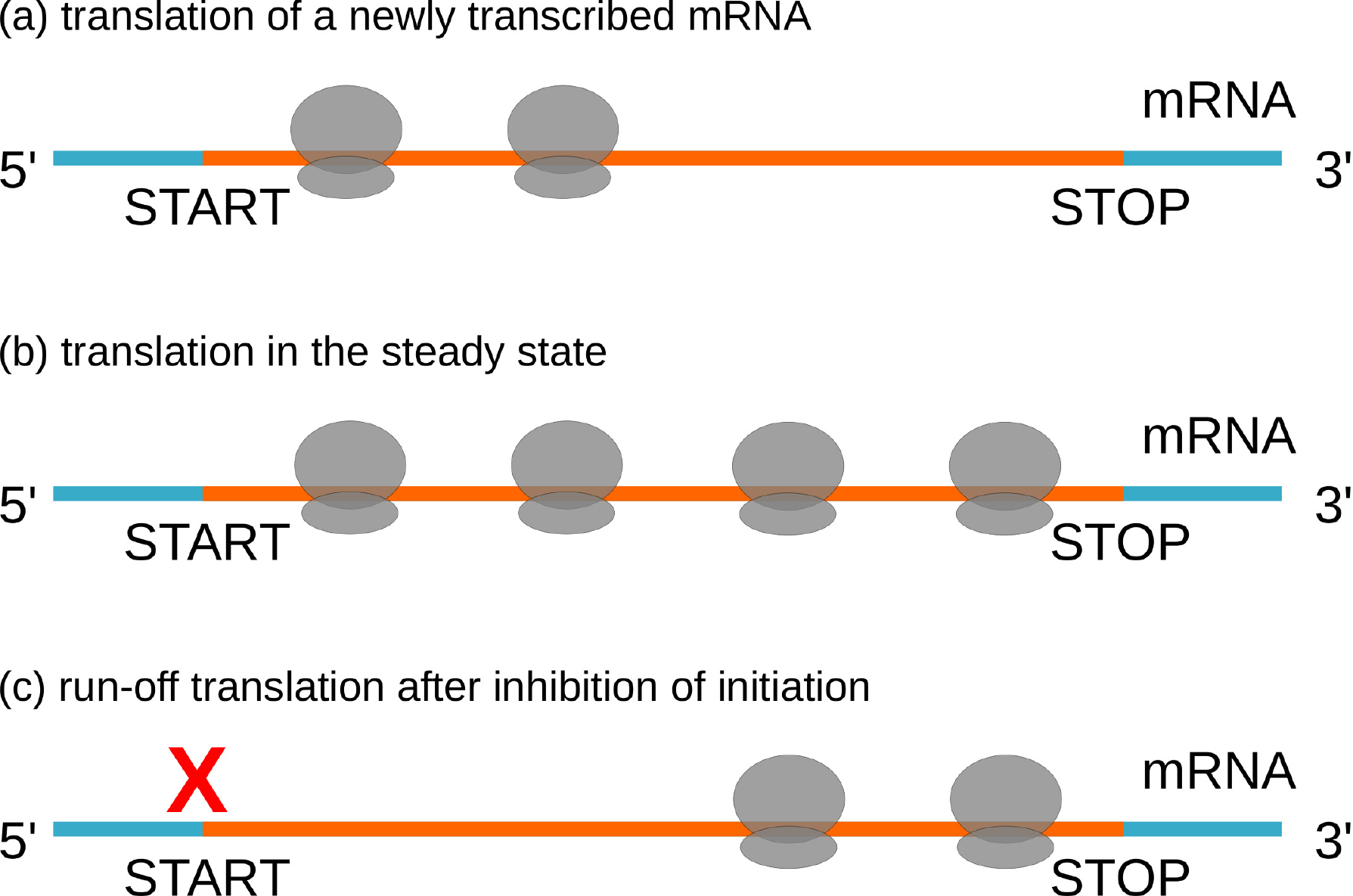}
	\caption{\label{fig1} Schematic of the three conditions that we study in the present work: (a) translation of a newly transcribed mRNA, (b) translation in the steady state and (c) run-off translation after inhibition of initiation.}
\end{figure}

In the present work, we model these recent experiments in the framework of the inhomogeneous TASEP that takes into account codon-specific elongation rates. Our goal is to understand the dynamics of translation under three conditions summarised in Fig.~\ref{fig1}:

\begin{itemize}
	\item[(a)] Translation of a newly transcribed mRNA, specifically the time evolution of the ribosome density and the time to reach the steady state (Fig.~\ref{fig1}(a))
	
	\item[(b)] Translation in the steady state, in particular the dynamics of individual (tagged) ribosomes, the time it takes a ribosome to translate a mRNA and the average speed of ribosomes (Fig.~\ref{fig1}(b))
	
	\item[(c)] Run-off translation after inhibition of  initiation (Fig,~\ref{fig1}(c)).
\end{itemize}
In each of these cases we develop analytic expressions that we benchmark against stochastic simulations for particular genes.

Mathematical models of translation are typically studied in the steady state and the goal is to compute the ribosome density and current. The novelty of our approach
is that we  consider the dynamics of translation under non-steady state conditions and also  the dynamics of individual (tagged) ribosomes in the steady state. The theory we present allows simple expressions for experimentally measurable quantities. Thus our work addresses a noteworthy gap that exists in the TASEP literature and provides a much needed framework for interpreting recent experiments that probe translation dynamics of individual ribosomes.

\section{\label{model} Kinetic model of \texorpdfstring{\MakeLowercase{m}RNA}{mRNA} translation}

\subsection{\label{definition} Definition of the model}

\begin{figure}[hbt]
	\centering
	\includegraphics[width=3.4in]{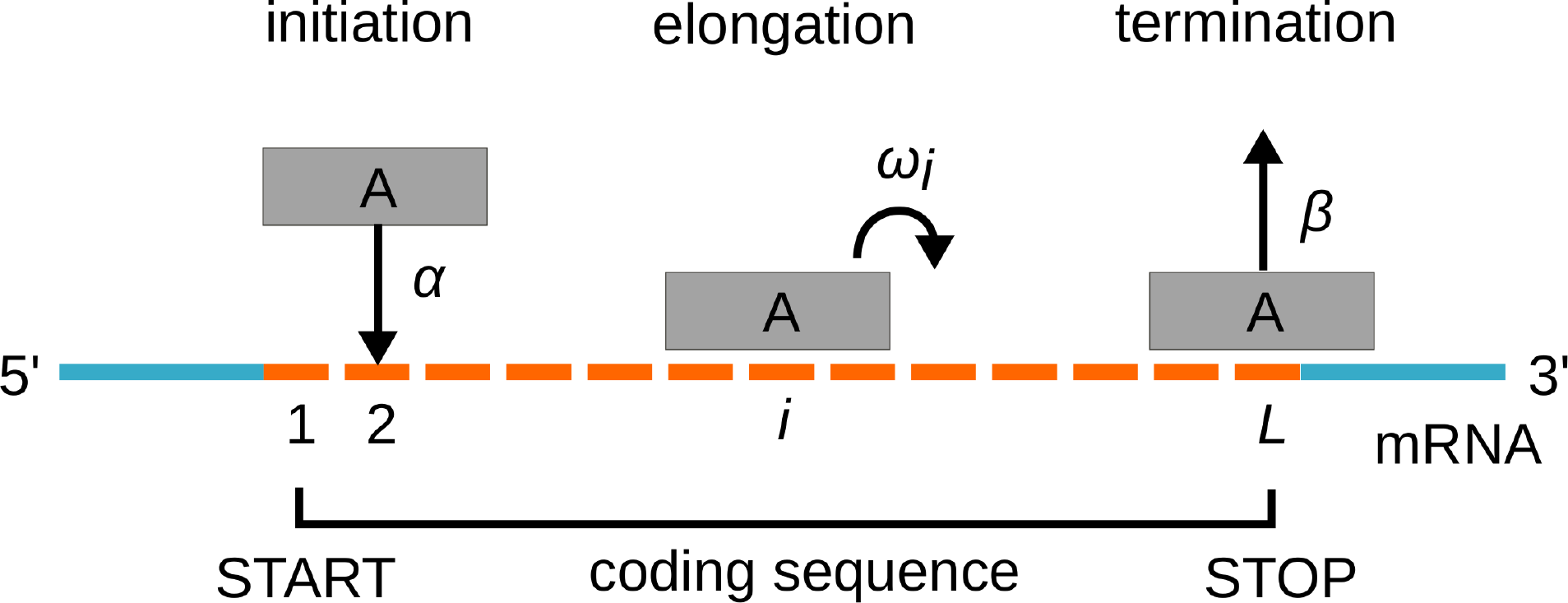}
	\caption{\label{fig2} Schematic of the TASEP with ribosomes of size $\ell=10$ and codon-dependent elongation rates $\omega_i$. Ribosomes occupy here $3$ codon positions only for demonstration.}
\end{figure}

We represent the mRNA molecule by one-dimensional lattice consisting of $L$ codons labelled from $1$ (start codon) to $L$ (stop codon) that code for $L-1$ amino acids (the stop codon does not code for an amino acid, see Fig.~\ref{fig2}). Each ribosome is a particle on the lattice occupying $\ell=10$ codons \cite{Ingolia2009}. A ribosome ``reads'' the mRNA sequence at its A-site, which is the site within the ribosome where the transfer RNA (tRNA) molecule delivers the correct amino acid. We assign an occupancy variable $\tau_i$ to each codon $i=2,\dots,L$ which takes value $1$ if the codon $i$ is occupied by the A-site and $0$ otherwise. Note that in this model site $1$ (start codon) is taken to be part of the initiation step. The occupancy vector $C=(\tau_2,\dots,\tau_L)$ keeps track of positions of all ribosomes on the lattice.

The model accounts for all three stages of translation: initiation, elongation and termination. Translation initiation involves a ribosome binding to the mRNA molecule and recognising the start codon. We model this process as a single step after which the A-site of the newly recruited ribosome is positioned at the second codon. The rate at which ribosomes attempt to initiate translation is denoted by $\alpha$ and is typically the slowest rate in the translation process under normal (physiological) conditions. The initiation is successful only if the codons $i=2,\dots,\ell+1$ are not occupied by an A-site of another ribosome. This step is summarised as:
\begin{equation}
\textrm{(initiation): $\tau_2=0\stackrel{\alpha}{\longrightarrow}1$ if $\tau_2=\dots=\tau_{\ell+1}=0$.}
\label{initiation}
\end{equation}
We note that our simplification of the initiation step accounts for both prokaryotic and eukaryotic translation initiation.

After initiation, a ribosome enters the elongation stage by receiving an amino acid from the corresponding tRNA and translocating to the next codon, provided there is no ribosome downstream blocking the move. Translation elongation at codon $i=2,\dots,L-1$ is modelled by the ribosome moving a one codon forward in a single step with codon-specific elongation rate $\omega_i$ (the inhomogeneous TASEP): the A-site moves from site $i$ to site $i+1$. This process is repeated at each codon until the ribosome A-site reaches the stop codon. This is the final stage of translation called termination during which the ribosome releases the polypeptide chain and unbinds from the mRNA. In the model, these steps are condensed into a single step that takes place at termination rate $\beta$. The elongation and termination stages are summarised as:
\begin{align}
\label{elongation}
&\textrm{(elongation): $\tau_i,\tau_{i+1}=1,0\stackrel{\omega_i}{\longrightarrow}0,1$ if $\tau_{i+\ell}=0$,}\nonumber\\
&\quad i=2,\dots,L-1,\\
\label{termination}
&\textrm{(termination): $\tau_{L}=1\stackrel{\beta}{\longrightarrow}0$.} 
\end{align}

Steps (\ref{initiation})-(\ref{termination}) constitute the original TASEP proposed in Ref.~\cite{MacDonald1968}. There are many other details of the translation process that may be added to the TASEP description that we do not consider here: multi-step elongation \cite{Fluitt2007,Basu2007,Zouridis2007,Ciandrini2010}, premature termination due to ribosome drop-off \cite{Gilchrist2006,Bonnin2017,Scott2019} and translation reinitiation due to mRNA circularisation  \cite{Chou2003,Gilchrist2006,Sharma2011,Marshall2014,Scott2019}, to name a few.

Often the problem with using more complex models is the lack of estimates for their kinetic parameters. In the case of mRNA circularisation (also known as the closed-loop model), the exact mechanism of how terminating ribosomes reinitiate at the start codon is not clear \cite{Vicens2018} and even less is known about the corresponding rate \cite{Gilchrist2006}. Previously, we analysed the TASEP with a simple reinitiation in which the terminating ribosomes initiates a new round of translation with a certain probability \cite{Scott2019}; this mechanism  was previously considered in Refs. \cite{Gilchrist2006,Marshall2014}. In particular, we showed that reinitiation has the same effect on ribosome density as increasing initiation rate in the model without reinitiation. Thus the conclusions drawn in the present work remain the same as long as the effective initiation rate is the rate-limiting step in translation.

In other cases such as ribosome drop-off the effect is small and can be ignored in the first approximation \cite{Scott2019}; the rate of ribosome drop-off in {\it E. coli} has been estimated to $10^{-3}$ s$^{-1}$ \cite{Sin2016,Bonnin2017}, which is about four orders of magnitude slower than the elongation rate. Multi-step elongation is an important addition to the basic model and even the two-step approximation of the elongation cycle consisting of tRNA delivery and translation can significantly alter the phase diagram of the TASEP \cite{Ciandrini2010}. 

Here, we limit our study to the basic model with codon-specific elongation rates mainly because dealing with the non-stationary TASEP--even the basic one--is a difficult problem. However, we note that none of the methods we use here are restricted to the basic TASEP.

\subsection{Master equation}

The TASEP is described by the probability $P(C,t)$ to find ribosomes in a configuration $C$ at time $t$, where $C$ records positions of all ribosomes on the lattice. The time evolution of $P(C,t)$ is governed by the master equation
\begin{align}
\label{master}
\frac{\partial P}{\partial t}&=\sum_{C'}W(C'\rightarrow C)P(C',t)\nonumber\\
&-\sum_{C'}W(C\rightarrow C')P(C,t).
\end{align}
where $C'\rightarrow C$ denotes transition from $C'$ to $C$ and $W(C'\rightarrow C)$ is the corresponding transition rate (initiation rate $\alpha$, elongation rates $\omega_2,\dots,\omega_{L-1}$ or termination rate $\beta$).

In the steady state $\partial P/\partial t=0$ and the master equation reduces to
\begin{equation}
\label{master_ss}
\sum_{C'}W(C'\rightarrow C)P^{*}(C')-\sum_{C'}W(C\rightarrow C')P^{*}(C)=0,
\end{equation}
where $P^{*}(C)$ is the steady-state distribution.

Traditionally, the late time dynamical behaviour of the homogeneous TASEP has been studied through the eigenvalue spectrum of (\ref{master}) \cite{deGier2005,Proeme2010} and current fluctuations \cite{Derrida1997,Gorissen2012}. The evolution from different initial conditions has been studied on the infinite system \cite{Schuetz1997,Imamura2007}. However these results are not of immediate utility in the translation context and for inhomogeneous TASEP. Therefore we take a more pragmatic approach.

\subsection{\label{parameters} Kinetic parameters}

We modelled translation of three genes, sodA from \textit{E. coli}, YAL020C from \textit{S. cerevisiae} and beta-actin from \textit{H. sapiens}. We used realistic kinetic parameters taken from the literature, which are summarised in Table \ref{tab1}.

\begin{table}[hbt]
	\caption{\label{tab1}Translation initiation and elongation rates for three genes that we studied.}
	\begin{ruledtabular}
		\begin{tabular}{lccc}
			Organism & Gene & Initiation rate & Elongation rates\\
			& & $\alpha$ [s$^{-1}$] &  $\omega_i$ [aa/s]\\
			\hline
			\textit{E. coli} & sodA & $1.0$ \cite{Gorochowski2019} & $4.7$---$56.6$ \cite{Rudolf2015}\\
			\textit{S. cerevisiae} & YAL020C & $0.08617$ \cite{Ciandrini2013} & $1.59$---$15.14$ \cite{Ciandrini2013}\\
			\textit{H. sapiens} & beta-actin & $0.0333$ \cite{Morisaki2016} & $10.0$ \cite{Morisaki2016}
		\end{tabular}
	\end{ruledtabular}
\end{table}

The genes were chosen based on the value of their initiation rate in order to represent different levels of ribosome traffic: sodA for fast ($\alpha=1$ s$^{-1}$), YAL020C for intermediate ($\alpha=0.08617$ s$^{-1}$) and beta-actin for slow ($\alpha=1/30$ s$^{-1}$) translation initiation. These rates were estimated from ribosome profiling \cite{Gorochowski2019}, polysome profiling \cite{Ciandrini2013} and fluorescence imaging experiments \cite{Morisaki2016}, respectively. 

Translation elongation rates for \textit{E. coli} and \textit{S. cerevisiae} genes were assumed to be codon-specific and were estimated from the concentrations of tRNA molecules delivering the corresponding amino acid \cite{Rudolf2015,Ciandrini2013}. For \textit{E. coli}, the rates were chosen at the doubling time of 96 min, which was the closest match to $85$ min reported in ribosome profiling experiments from which the initiation rates were inferred \cite{Gorochowski2019}. For beta-actin gene we used an average elongation rate of $10$ aa/s inferred from fluorescence imaging experiments \cite{Morisaki2016}.

Translation termination is typically fast, but the specific data on the rates are lacking. In our model we assume that ribosomes terminate immediately after they reach the stop codon so that effectively $\beta \gg \alpha$, which is a common assumption in modelling translation \cite{Shah2013}. A fast termination is consistent with the results of ribosome profiling experiments showing an increased ribosome activity at the stop codon but  without ribosome queues \cite{Ingolia2011,Weinberg2016}. Indeed, recent estimates of termination rates from ribosome profiling data in {\it S. cerevisiae} suggest that the termination rate is an order of magnitude larger than the initiation rate \cite{DaoDuc2018,SzavitsNossan2019}. Although that is far from termination being instantaneous as in our model, setting a finite value of the termination rate does not change our conclusions as long as it is larger than the initiation rate.

\subsection{\label{density-current} Ribosome density and current}

Ribosome density $\rho_i(t)$ determines how likely it is to find a ribosome at site $2\leq i\leq L$ at time $t$ and is defined as
\begin{equation}\label{rho_i_def}
\rho_i(t)=P(\tau_i(t)=1)=\sum_{C}\tau_i(C)P(C,t).
\end{equation}
The average density $\rho(t)$ is equal to the average number of ribosomes at time $t$ divided by $L-1$,
\begin{equation}
\label{rho_t_def}
\rho(t)=\frac{1}{L-1}\sum_{i=2}^{L}\rho_i(t).
\end{equation}
The steady-state densities $\rho^{*}_{i}$ and $\rho^{*}$ are defined as above with $P(C,t)$ replaced by the steady-state distribution $P^{*}(C)$. Translation is a nonequilibrium process since there is always a flow of ribosomes. In the nonequilibrium steady state, the current of ribosomes $J^{*}$ is constant across the mRNA and is equal to
\begin{subequations}
\begin{align}
J^{*}&=\alpha P^{*}(\tau_2=\dots\tau_{\ell+1}=0)\\
&=\omega_i P^{*}(\tau_i=1,\tau_{i+\ell}=0),\; i=2,\dots,L-\ell \label{Jdef}\\
&=\omega_i P^{*}(\tau_i=1),\; i=L-\ell+1,\dots,L\\
&=\beta P^{*}(\tau_L=1).
\end{align}
\end{subequations}
It is useful to write $J^{*}$ in a slightly different way by noting that we may write the joint probability $P^{*}(\tau_i=1,\tau_{i+\ell}=0)$ as $P^{*}(\tau_i=1,\tau_{i+\ell}=0)=\rho_i P^{*}(\tau_{i+\ell}=0\vert \tau_i=1)$, where $P^{*}(\tau_{i+\ell}=0\vert \tau_i=1)$ is the conditional probability that codon $i+\ell$ is empty, given that the codon $i$ is occupied. $P^{*}(\tau_{i+\ell}=0\vert \tau_i=1)$ measures the efficiency of elongation at codon $i$ and takes values between $0$ and $1$ depending on the level of ribosome traffic. On the other hand, $P(\tau_2=\dots=\tau_{\ell+1}=0)$ measures how likely is for the initiation to be successful depending on the traffic around the start codon. We will refer to $P^{*}(\tau_{i+\ell}=0\vert \tau_i=1)$ and $P(\tau_2=\dots=\tau_{\ell+1}=0)$ as the translation elongation efficiency (TEE$_i$) and translation initiation efficiency (TIE), respectively \cite{SzavitsNossan2019}. Using these definitions, the current $J^{*}$ can be written as
\begin{equation}
\label{current_TIE_TEE}
J^{*}=\alpha\cdot\textrm{TIE}=\omega_i\rho_{i}^{*}\textrm{TEE}_i.
\end{equation}
We note that we set  $\textrm{TEE}_i=1$ for $i=L-\ell+1,\dots,L-1$.

Throughout this paper we assume that the steady-state densities $\rho_i^*$ and current $J^*$ are known; we obtain these from stochastic simulations of the model. Alternatively, one can compute $\rho_i^*$ and $J^*$ using the mean-field theory \cite{ErdmannPham2020} and the power series method \cite{Szavits2018,Scott2019}.

\begin{figure*}[htb]
	\centering
	\includegraphics[height=1.51in]{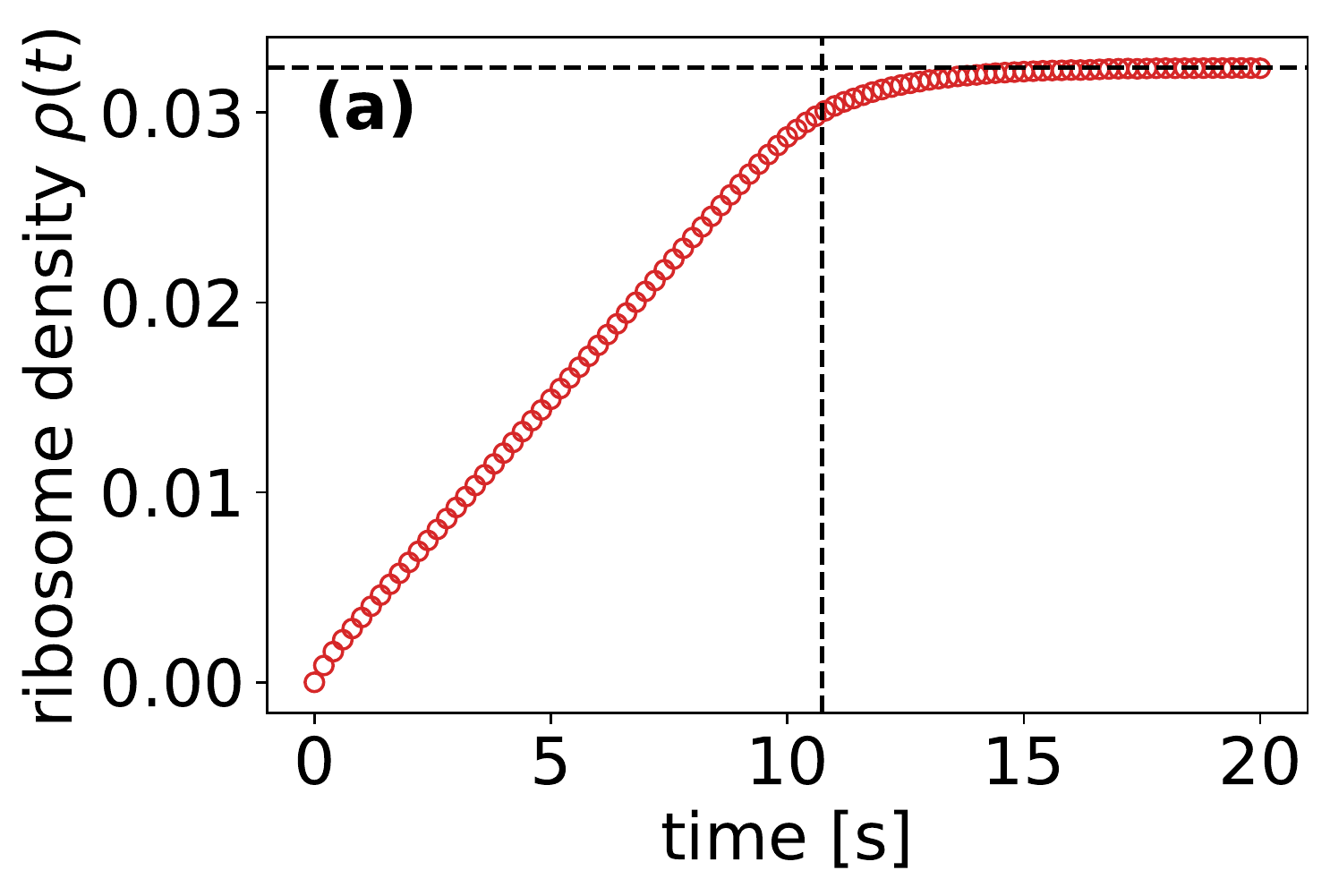}
	\includegraphics[height=1.51in]{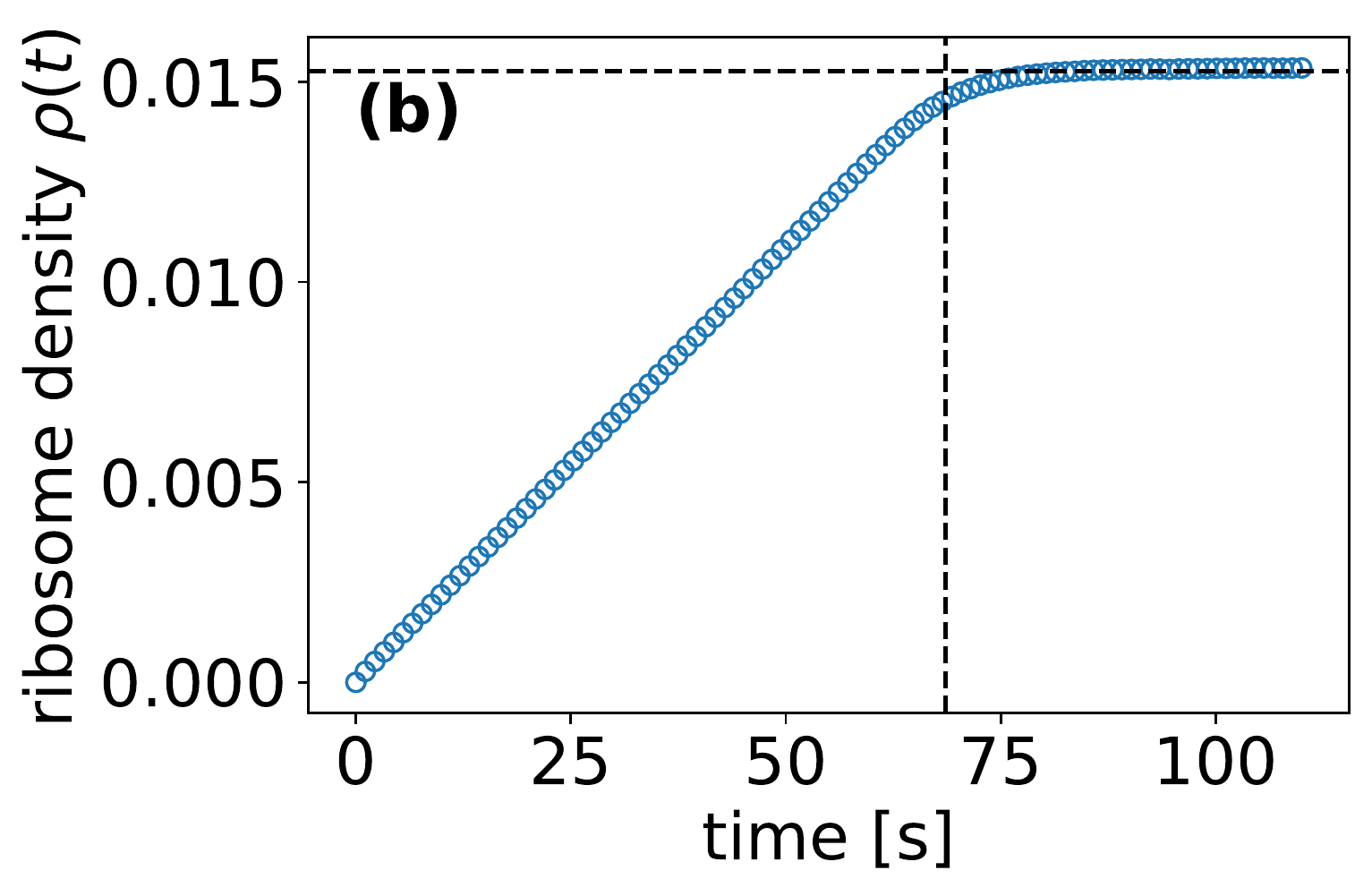}
	\includegraphics[height=1.51in]{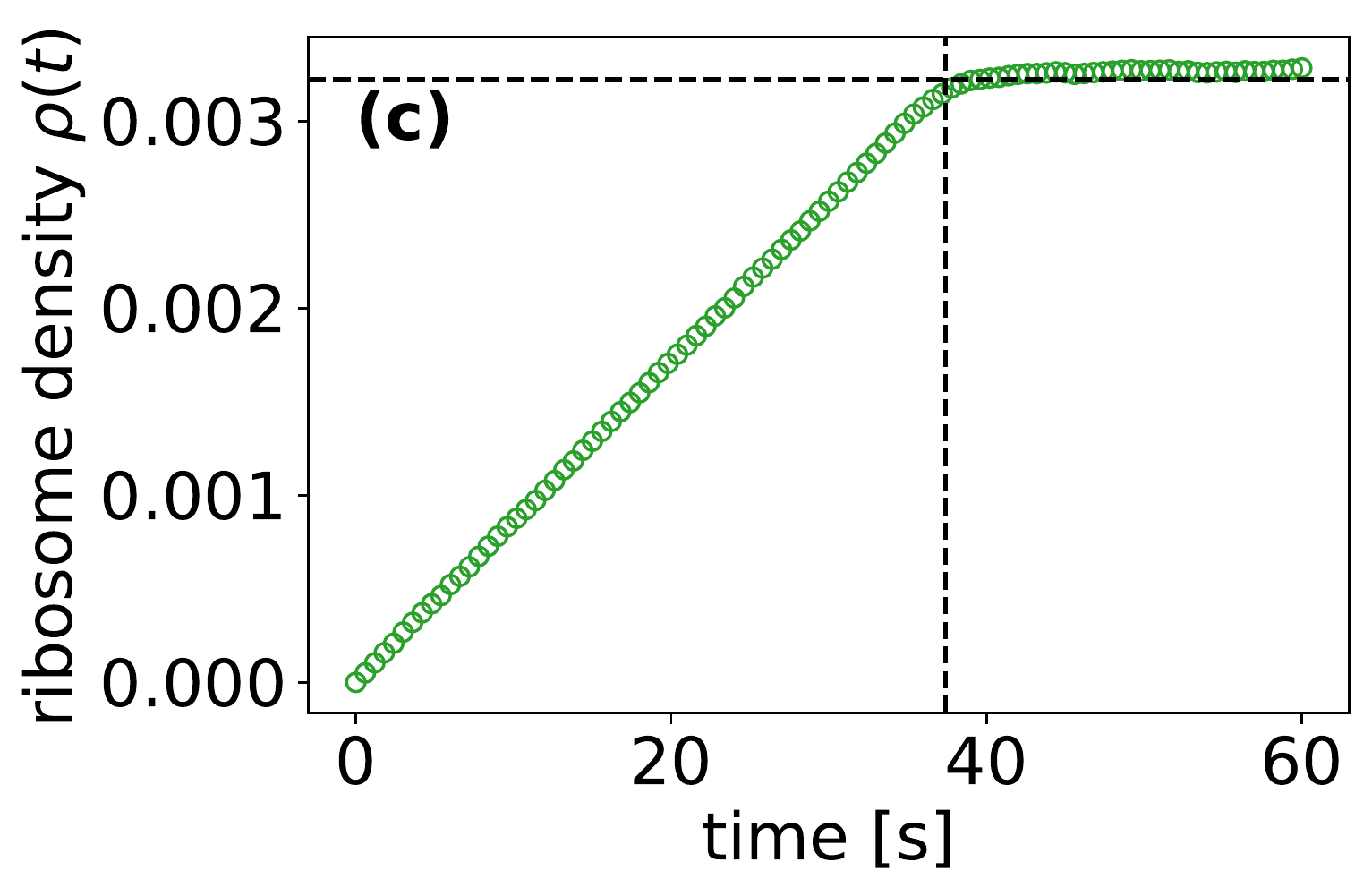}
	\caption{\label{fig3} Time evolution of the total ribosome density $\rho(t)$ for genes (a) sodA of \textit{E. coli}, (b) YAL020C of \textit{S. cerevisiae} and (c) beta-actin of \textit{H. sapiens}, obtained from stochastic simulations averaged over $10^6$ independent runs. The corresponding initiation rates are $1$, $0.08617$ and $1/30$ initiations/s. Vertical dashed lines mark the average translation time $\langle T\rangle$ of the first round of translation.}
\end{figure*}
	
\section{Translation of a newly transcribed \texorpdfstring{\MakeLowercase{m}RNA}{mRNA}}

We first consider time evolution of the total ribosome density $\rho(t)$ from a newly transcribed mRNA. We track ribosomes as they translate the mRNA leading to an increase in $\rho(t)$ over time. Eventually the system settles in the steady state and $\rho(t)\approx \rho^{*}$ for $t$ larger than some characteristic time $t_{0}$. Our goal is to understand how $t_0$ depends on the model's parameters.

\subsection{Translation time of the first round of translation} 

In Fig.~\ref{fig3} we plot the time evolution of the total ribosome density obtained by stochastic simulations using realistic kinetic parameters for genes sodA \textit{E. coli}, YAL020C of \textit{S. cerevisiae} and beta-actin of \textit{H. sapiens}. Despite gene-specific differences in the number of codons and the kinetic parameters, all three genes display similar time evolution consisting of a linear increase followed by a plateau at the corresponding value of the steady-state density $\rho^{*}$. We further observe that the time $t_0$ to reach the steady state is very close to the average translation time $\langle T\rangle$ of the first round of translation (vertical dashed lines in Fig. \ref{fig3}). These observations are consistent with fluorescence imaging experiments of newly transcribed mRNAs that show a linear increase in the fluorescence signal until the end of the first round of translation \cite{Yan2016}. We note that the linear increase in the ribosome density has been observed before in the homogeneous TASEP in which the elongation rates are constant along the transcript \cite{Nagar2011}.

A sharp transition from the linear increase to the plateau is indicative of translation that is rate-limited by initiation. Indeed, the rates of initiation of all three genes in Fig.~\ref{fig3} are smaller than the elongation rates of their individual codons. The best agreement between $\langle T\rangle$ (the end of the linear increase) and $t_{0}$ (the beginning of the plateau) is found for beta-actin gene which initiates at the rate of $\alpha=1/30$ initiations/s. The least agreement is found for sodA gene which initiates $30$ times faster than $\beta$ actin. In that case the linear increase which ends after $\langle T\rangle$ is followed by a slower nonlinear increase towards the steady state value $\rho^*$. The nonlinear regime is characteristic of high initiation rates, which lead to increased ribosome traffic and slower relaxation dynamics.

Based on these observations we use the translation time of the pioneering round as a proxy for the time to reach the steady state. The translation time $T$ is equal to the sum of dwell times $e_i$ at each codon $i=2,\dots,L$
\begin{equation}
\label{T}
T=e_2+e_3+\dots+e_{L}.
\end{equation}
Because the pioneering ribosome moves across an empty mRNA, the probability density function (PDF) of $e_i$ is simply
\begin{equation}
p_k(e_k)=\omega_k\textrm{exp}(-\omega_k e_k),\quad k=2,\dots,L.
\end{equation}
The sum of exponential random variables in Eq.~(\ref{T}) follows the hypoexponential distribution (see Appendix \ref{appendix_a} for details). The probability density function (PDF) $p(T)$ and the cumulative distribution function (CDF) $P(T)$ are known explicitly and are given by
\begin{subequations}
	\begin{align}
	\label{pT1}
	& p(T)=\sum_{k=2}^{L}\omega_k\textrm{e}^{-\omega_k T}\left(\prod_{j=2, j\neq k}^{L}\frac{\omega_j}{\omega_j-\omega_k}\right),\\
	\label{PT1}
	& P(T)=1-\sum_{k=2}^{L}\textrm{e}^{-\omega_k T}\left(\prod_{j=2, j\neq k}^{L}\frac{\omega_j}{\omega_j-\omega_k}\right).
	\end{align}
\end{subequations}
When all elongation rates $\omega_i=\omega$, the distribution reduces to the Erlang distribution. The mean and variance of $T$ are
\begin{equation}
\label{t1_s1}
\langle T\rangle=\sum_{j=2}^{L}\frac{1}{\omega_j},\quad\langle T^2\rangle-\langle T\rangle^2=\sum_{j=2}^{L}\frac{1}{\omega_{j}^{2}}.
\end{equation}

\begin{figure}[hbt]
	\centering\includegraphics[width=3.4in]{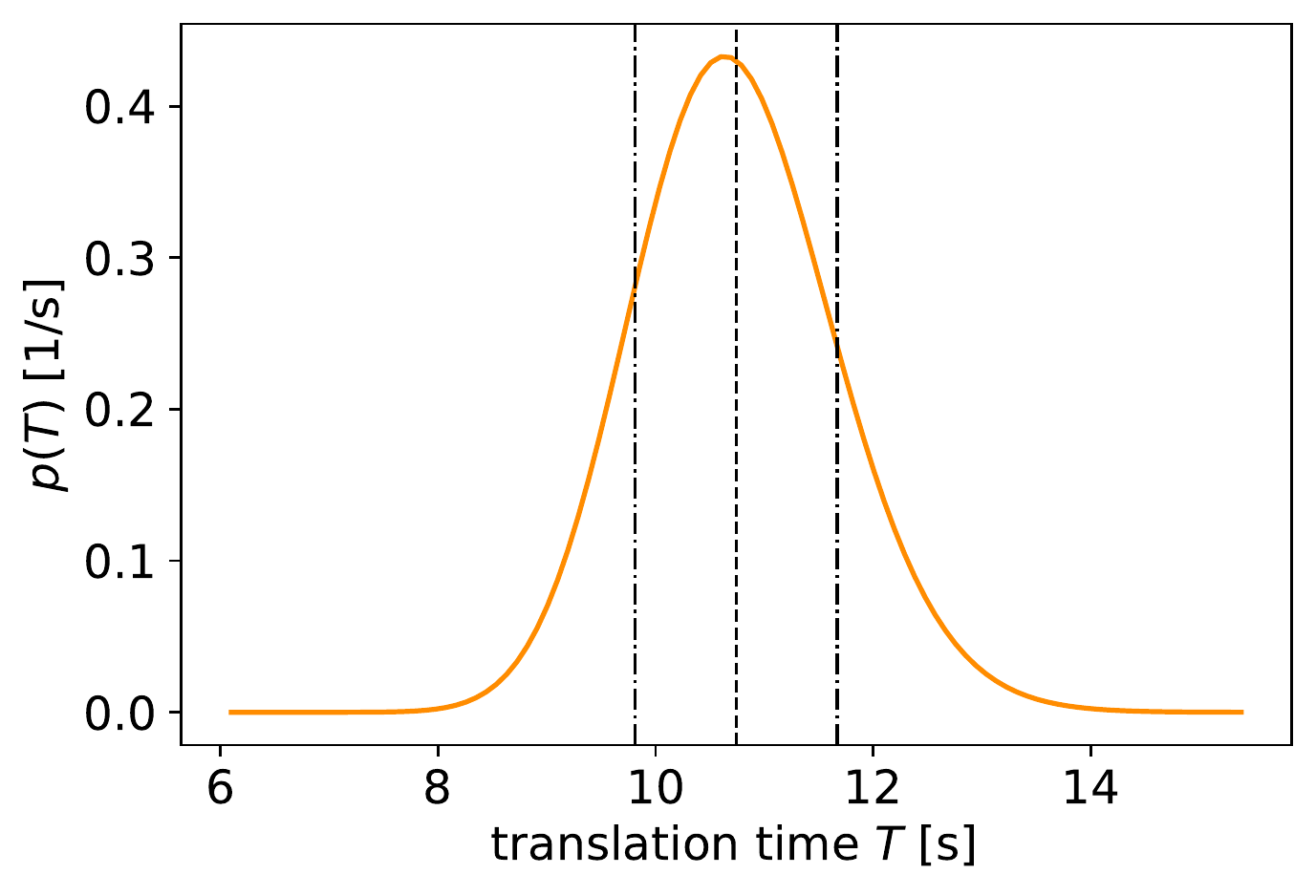}
	\caption{\label{fig4} Probability density function $p(T)$ for the translation time $T$ of the first, pioneering round of translation of gene sodA. Vertical dashed line is at the mean value $\langle T\rangle$ and the dashed-dotted lines are at $\langle T\rangle\pm(\langle T^2\rangle-\langle T\rangle^2)^{1/2}$.}
\end{figure}

The probability density function for the translation time $T$ of sodA gene is plotted in Fig. \ref{fig4}. We note that the expression for $p(T)$ in Eq.~(\ref{pT1}) can produce significant rounding errors due to extremely small values of the products in the sum. Instead, we used an alternative expression for $p(T)$ that includes a matrix exponential (see Appendix \ref{appendix_a} for details). The matrix exponential was then computed using \texttt{linalg.expm} algorithm from the SciPy library. When the number of codons is large, the distribution can be approximated by a Gaussian distribution, which is due to the central limit theorem for independent but not identically distributed random variables \cite{Feller1968}.

\begin{figure*}[htb]
	\centering
	\includegraphics[height=1.53in]{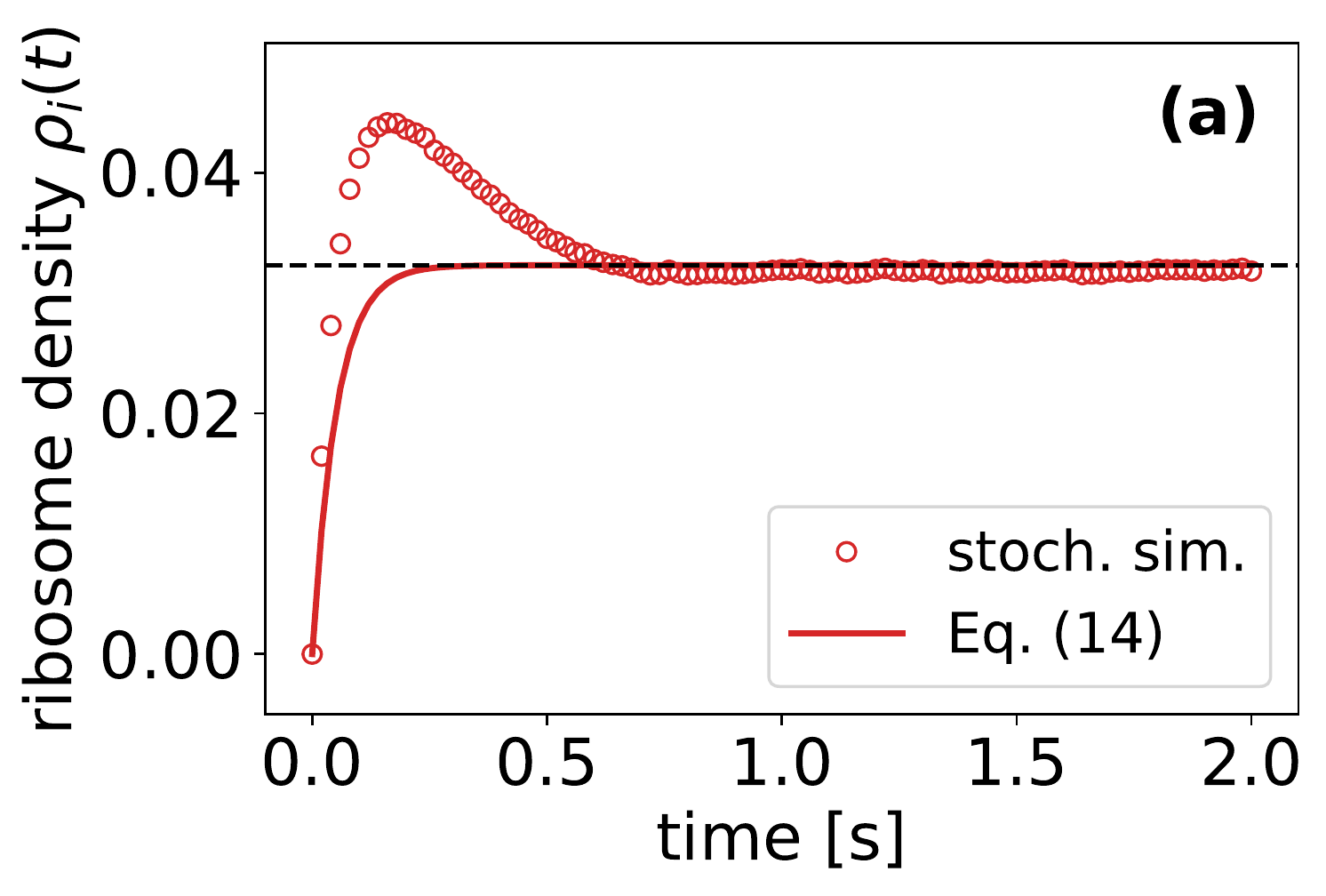}
	\includegraphics[height=1.53in]{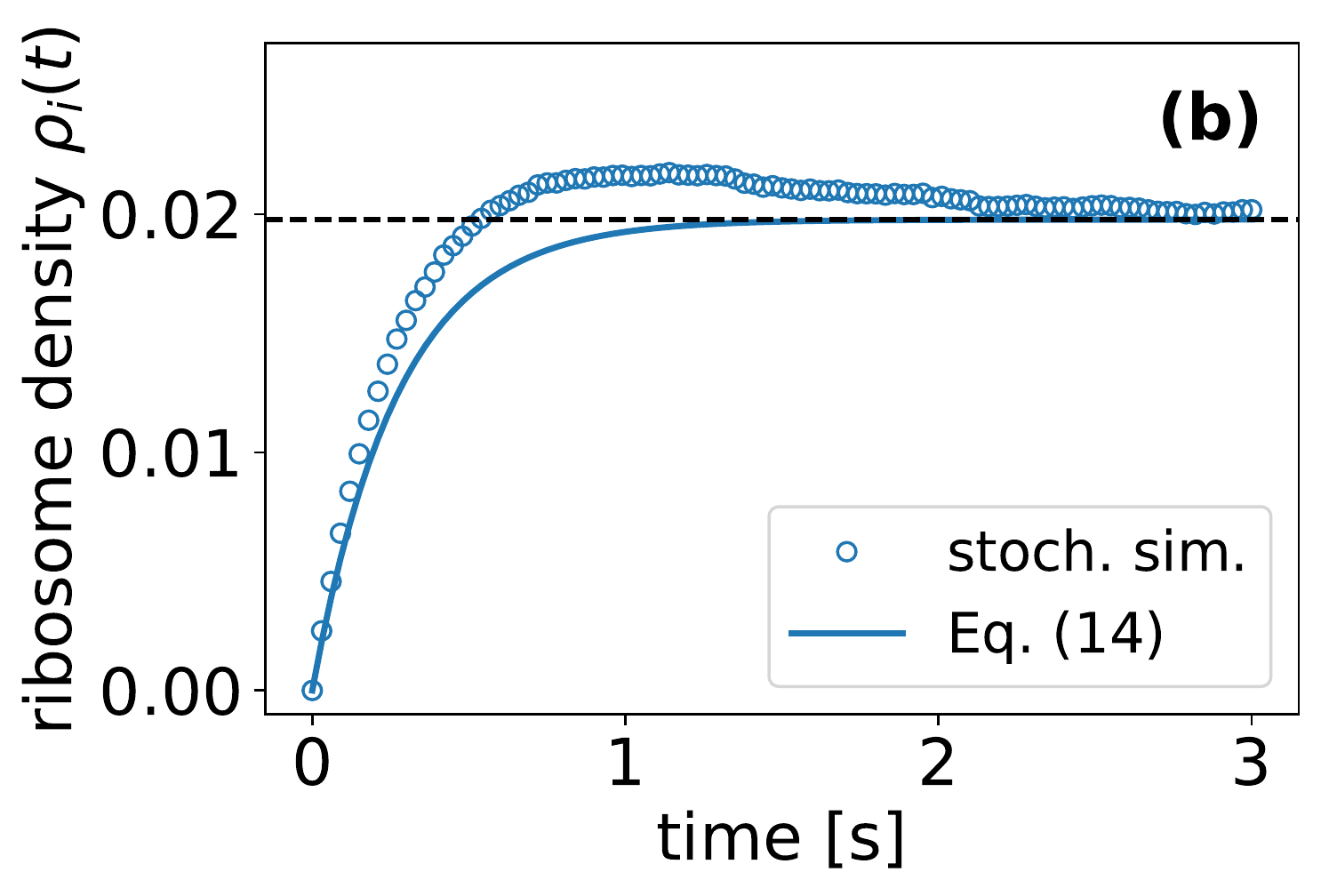}
	\includegraphics[height=1.53in]{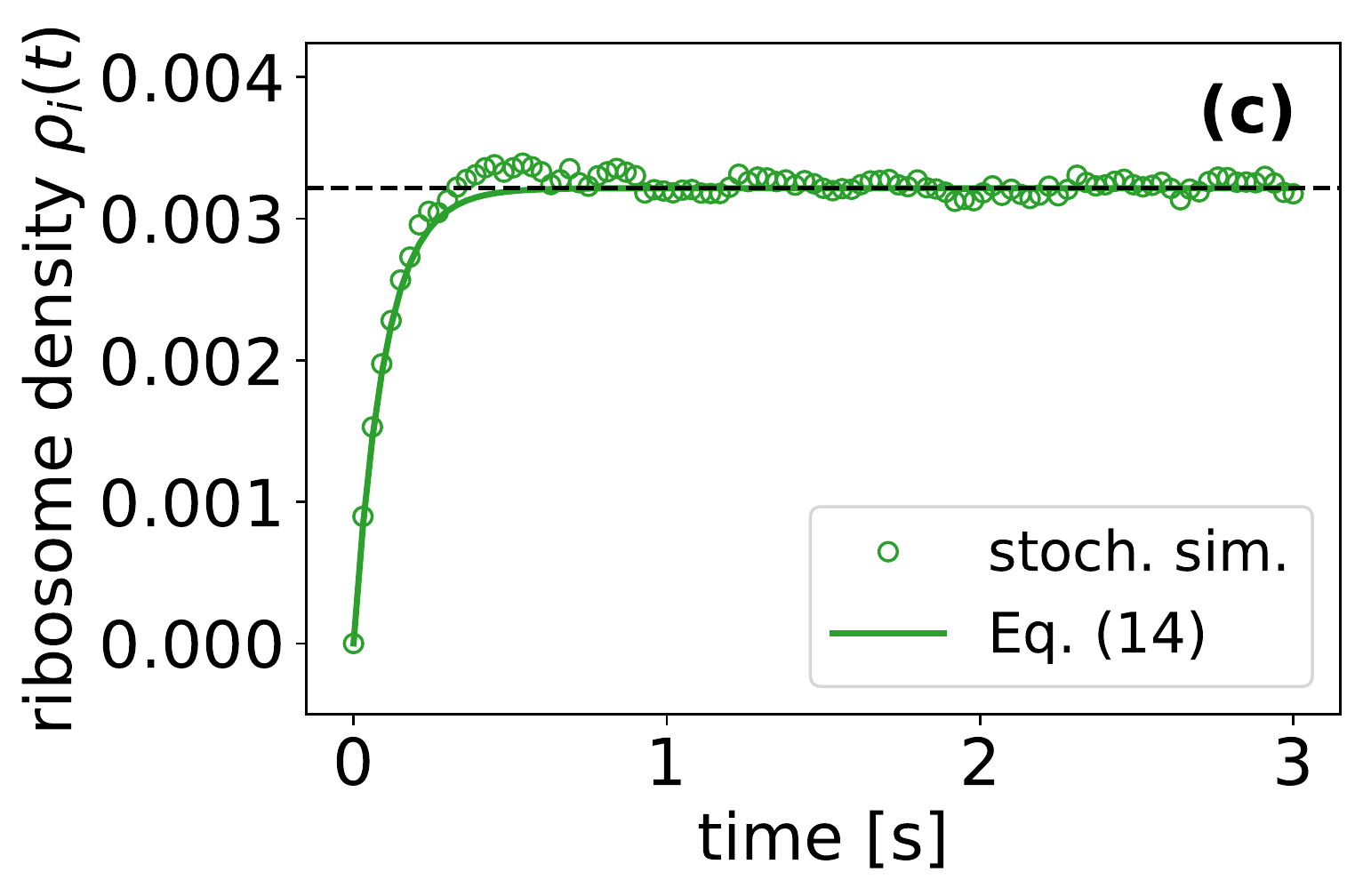}
	\includegraphics[height=1.53in]{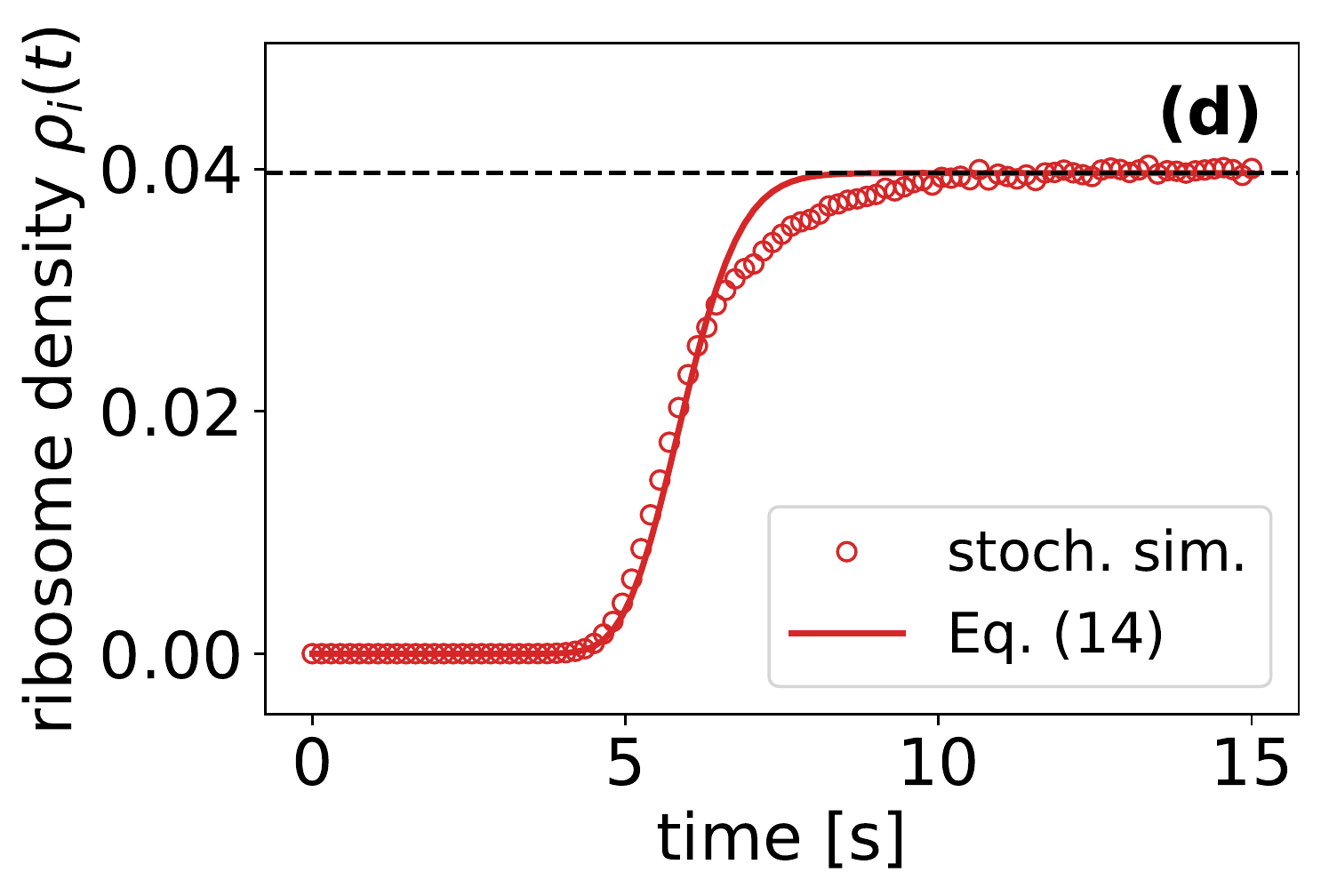}
	\includegraphics[height=1.53in]{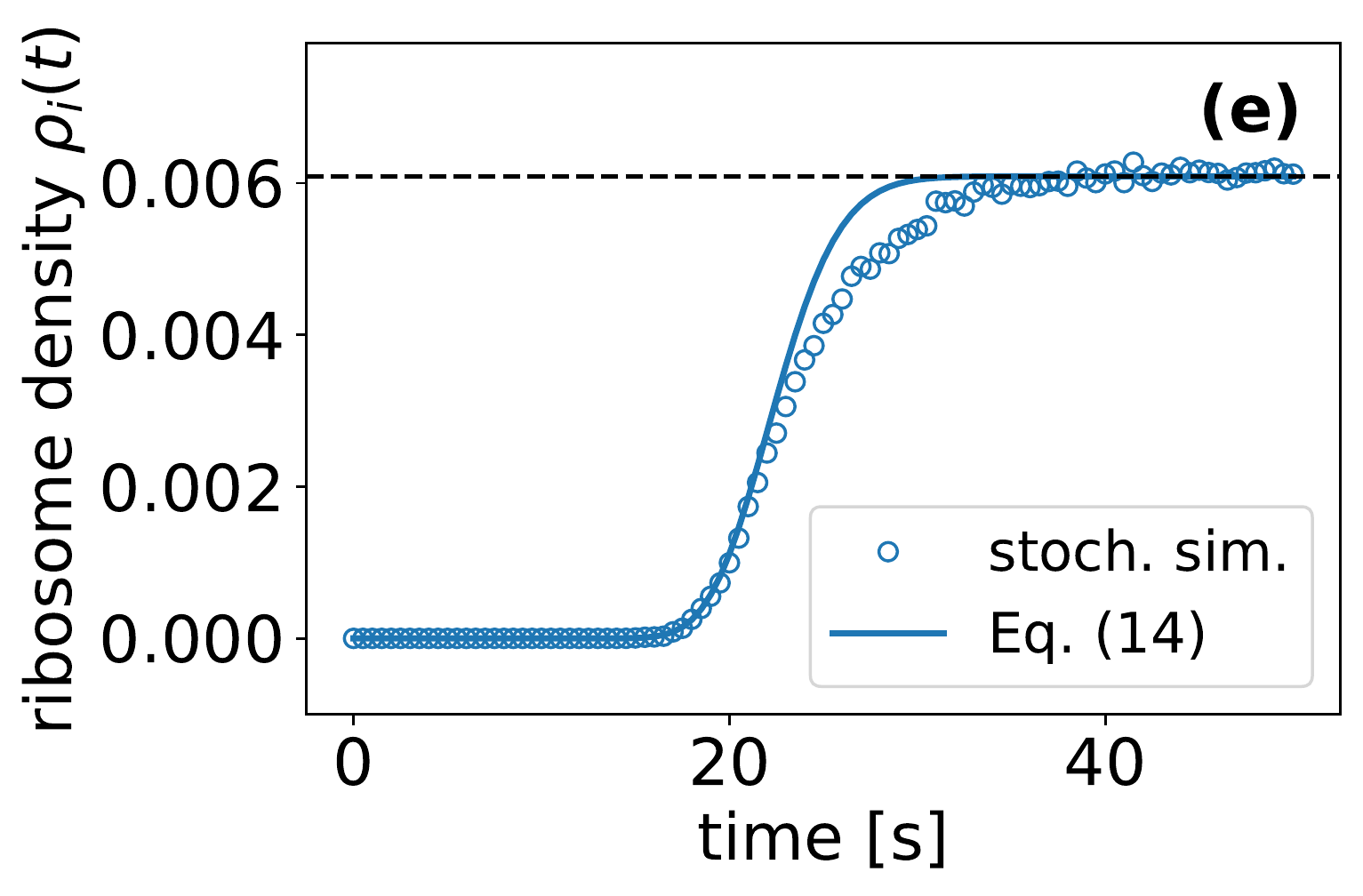}
	\includegraphics[height=1.53in]{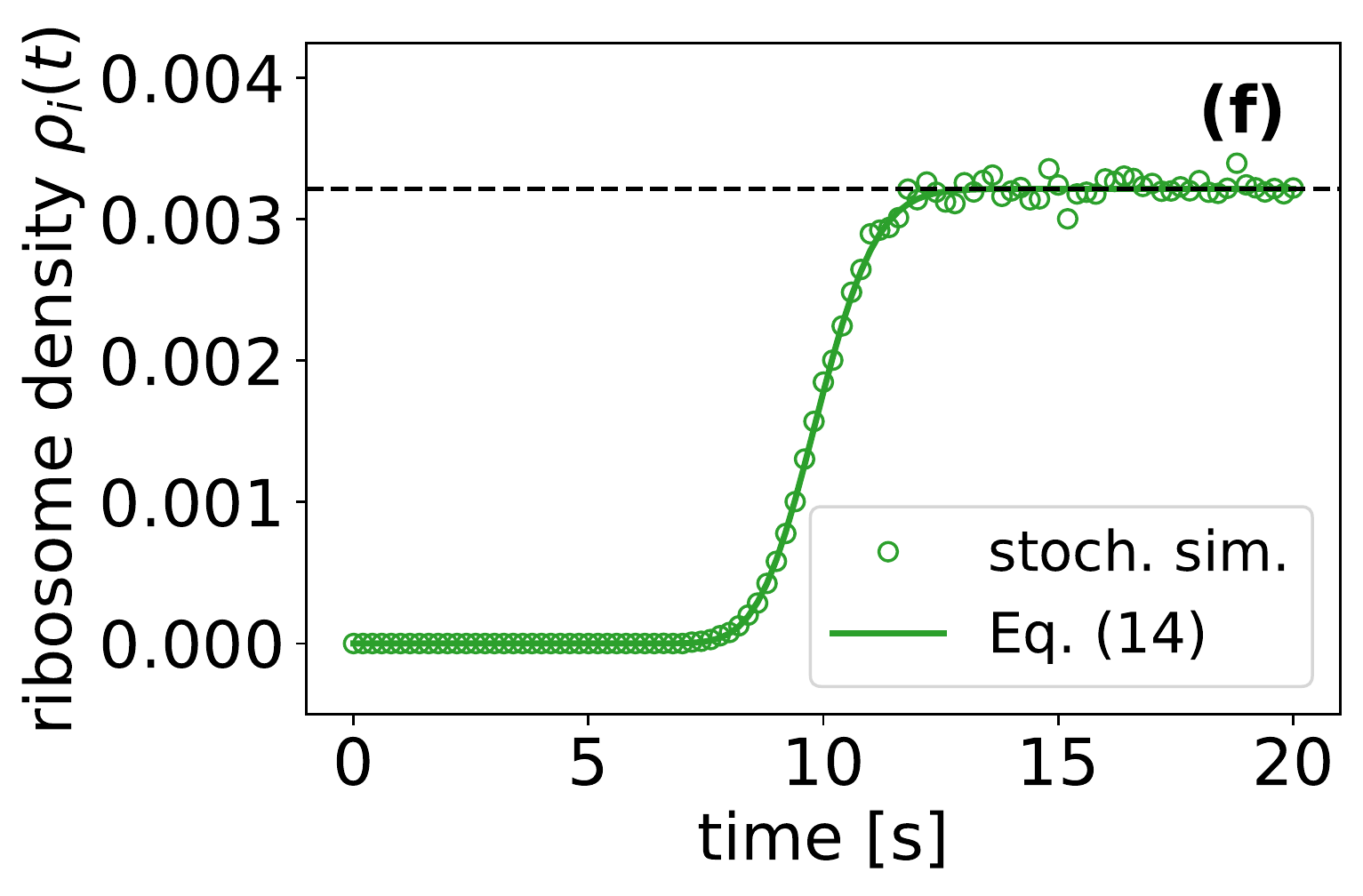}
	\caption{\label{fig5} (a)-(f) Time evolution of the ribosome density $\rho_i(t)$ for genes sodA (left column), YAL020C (middle column) and beta-actin (right column), at two codon positions, $i=2$ (top row) and $i=100$ (bottom row). Symbols are from stochastic simulations averaged over $10^6$ independent runs. Dashed lines were computed using Eq.~(\ref{rho_i_1}). Horizontal dashed line is the steady-state density $\rho_{i}^{*}$.}
\end{figure*}

\subsection{Time evolution of the ribosome density \texorpdfstring{$\rho_i(t)$}{rhoi(t)} and the total ribosome density \texorpdfstring{$\rho(t)$}{rho(t)}}

So far we have seen that when translation is rate-limited by initiation, the steady state is reached shortly after the end of the pioneering round of translation. We now extend this result to any codon position $i$ and assume that the steady-state density $\rho_{i}^{*}$ is reached as soon as the pioneering ribosome leaves the site $i$. Under this assumption,
\begin{equation}
\label{rho_i_1}
\rho_i(t)\approx\rho_{i}^{*}P\left(\sum_{j=2}^{i}e_j\leq t\right),
\end{equation}
where $P$ is the same as in Eq. (\ref{PT1}) except that $L$ is replaced by $i$. Approximation \ref{rho_i_1}  simply expresses $\rho_i(t)$ as an average over two values $0$ or $\rho_{i}^{*}$ depending on the random position of the pioneering ribosome. In Fig.~\ref{fig5} we plot time evolution of the ribosome density $\rho_i(t)$ at two codon positions, $i=2$ and $i=100$. The simple expression in Eq.~(\ref{rho_i_1}) agrees well with the results of stochastic simulation for beta-actin gene, but less so for genes YAL020C and sodA which have $2.5$ and $30$ times faster initiation rate, respectively. The overshoot of the density noticeable in Fig.~\ref{fig5}(a)-(c) is due to the first ribosome entering the lattice. The density then drops down as the first ribosome moves away, but rises again due to the next ribosome. After few of these oscillations, the density eventually flattens due to the steady flux of ribosomes. In general, we expect Eq.~(\ref{rho_i_1}) to be accurate for small $\alpha$ such that ribosome collisions are rare.

\begin{figure}[hbt]
	\centering
	\includegraphics[width=3.4in]{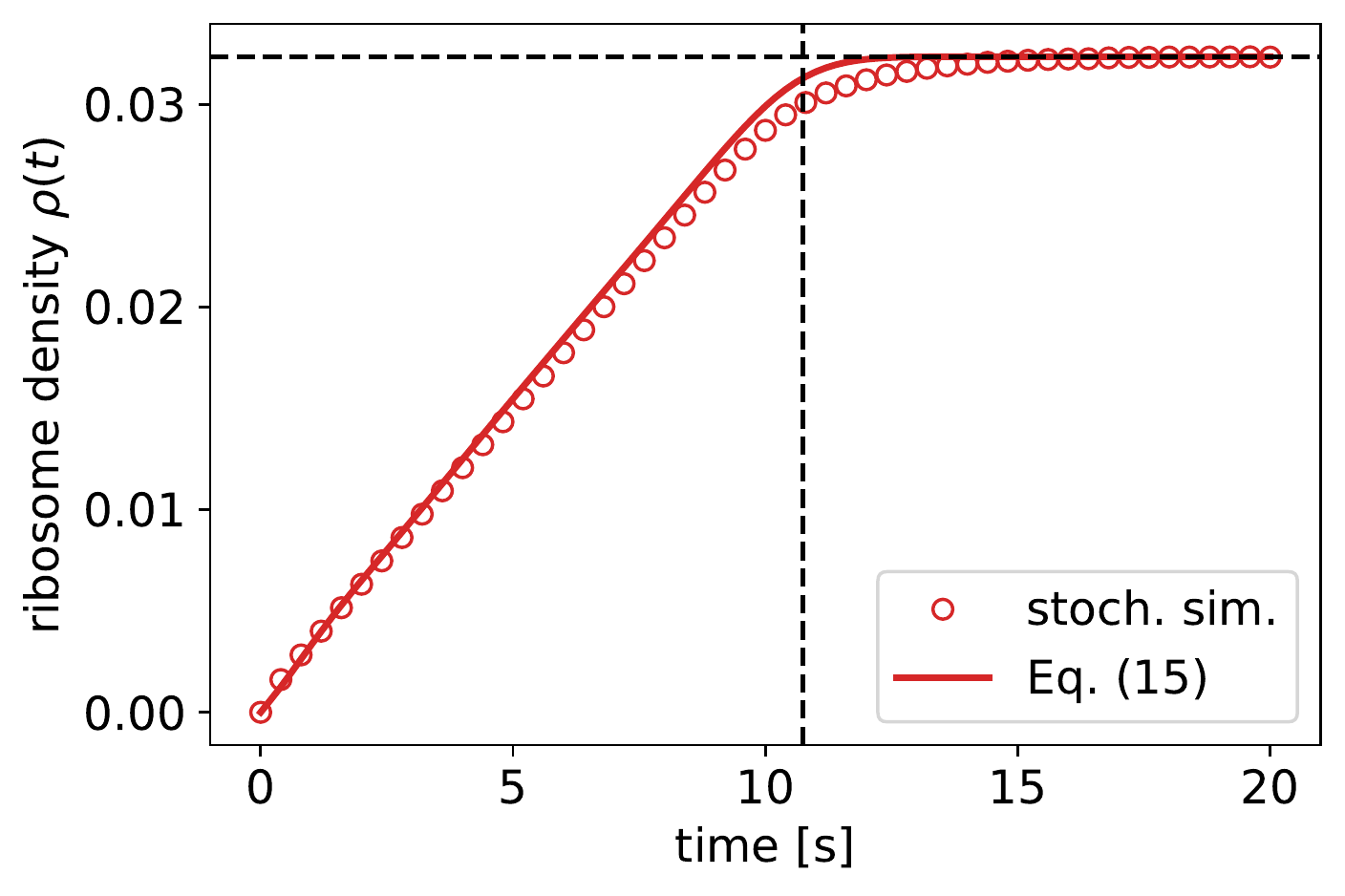}
	\caption{\label{fig6} Time evolution of the ribosome density $\rho(t)$ for gene sodA of \textit{E. coli}. Symbols are from stochastic simulations averaged over $10^6$ independent runs. Solid line was computed using Eq.~(\ref{rho_t}). Horizontal dashed line is the steady-state density $\rho^{*}$.}
\end{figure}

The time evolution of the total  density $\rho(t)$ is obtained by inserting Eq.~(\ref{rho_i_1}) into (\ref{rho_t_def}), 
\begin{equation}
\label{rho_t}
\rho(t)\approx\frac{1}{L-1}\sum_{i=2}^{L}\rho_{i}^{*}P\left(\sum_{j=2}^{i}e_j\leq t\right).
\end{equation}
This expression reproduces the linear increase followed by a plateau observed in Fig.~\ref{fig3}, which we demonstrate for sodA gene in Fig.~\ref{fig6}.

\section{Translation in the steady state}

In this Section we want to understand the dynamics of individual (tagged) ribosomes after the system has settled in the steady state. We tag a ribosome that initiated translation at some reference time $t=0$ and track its position $X(t)$ along the mRNA. We denote by $T^*$ the (stochastic) translation time it takes the ribosome to move across the mRNA and terminate at the stop codon. Our goal is to find the distribution of $X(t)$ and $T^*$.

\subsection{The effective medium approximation}

The probability that the ribosome at codon position $X(t)=n$ moves to $X(t+\Delta t)=n+1$ during a small time interval $\Delta t$ is given by
\begin{align}
& P(X(t+\Delta t)=n+1\vert X(t)=n)  = (\omega_n \Delta t)\nonumber\\
& \qquad \times P(\tau_{n+\ell}(t) =0\vert X(t) =n).
\label{PX}
\end{align}
Probabilities involving the tagged particle position $X(t)$ are more complicated objects than the particle density and exact results are rare \cite{Spitzer1970,Kipnis1986,Imamura2007}. To make progress we approximate the right hand side of (\ref{PX}) with  the steady-state conditional probability $P^*(\tau_{n+\ell}=0\vert\tau_n=1)$ which in turn may be written as $\displaystyle P^*(\tau_{n+\ell}=0\vert\tau_n=1)= P^*(\tau_{n+\ell}=0\vert\tau_n=1)/P^{*}(\tau_n=1)$ or using definitions of current and density (\ref{Jdef} and \ref{rho_i_def}) in the steady state
\begin{equation}
\label{EMA}
 P(X(t+\Delta t)=n+1\vert X(t)=n)
=\frac{J^{*}}{\rho_{n}^{*}}\Delta t\;.
\end{equation}
To arrive at this approximation, we have assumed that the process as seen by the tagged ribosome has the same steady-state probability distribution $P^{*}(C)$ as the original process.  

The probability that the tagged particle stays at codon $n$ in the small time interval $\Delta t$ is $1-J^{*}/\rho_{n}^{*}\Delta t$, which means that the dwell time $e_{n}^{*}$ of the tagged ribosome at codon $n$ follows an exponential distribution with the effective rate $\lambda_n=J^{*}/\rho_{n}^{*}$,
\begin{equation}
\label{lambda}
p_n(e_{n}^{*})=\lambda_n\textrm{e}^{-\lambda_n e_{n}^{*}},\quad \lambda_n=\frac{J^*}{\rho_{i}^{*}}.
\end{equation}
We call Eq.~(\ref{EMA}) the \emph{effective medium approximation}, because it reduces the dynamics of a tagged ribosome in the dynamic environment made of other ribosomes to a continuous-time random walk of a single ribosome with effective rates $\lambda_n$. Using the definition of $J^{*}$ from Eq.~(\ref{current_TIE_TEE}), the effective rate $\lambda_n$ becomes $\lambda_n=\omega_{n}\textrm{TEE}_n$, where TEE$_n$ is the translation elongation efficiency. TEE$_n$, which takes values between $0$ and $1$, is a simple measure of ribosome traffic that allows us to understand how the dynamics of a single ribosome is affected by other ribosomes on the mRNA.  

Interestingly, the effective medium approximation becomes \emph{exact} in the infinite TASEP with particles of size $\ell=1$ and homogeneous elongation rates $\omega$, provided the system is initially in the steady state in which $\rho_{i}^{*}=\rho^{*}$ (the Bernoulli measure). In that case $J^{*}=\omega\rho^{*}(1-\rho^{*})$ and the position $X(t)$ of the tagged particle follows the Poisson distribution with rate $\omega(1-\rho^{*})t$ \cite{Spitzer1970}.

\begin{figure*}[htb]
	\centering
	\includegraphics[height=1.55in]{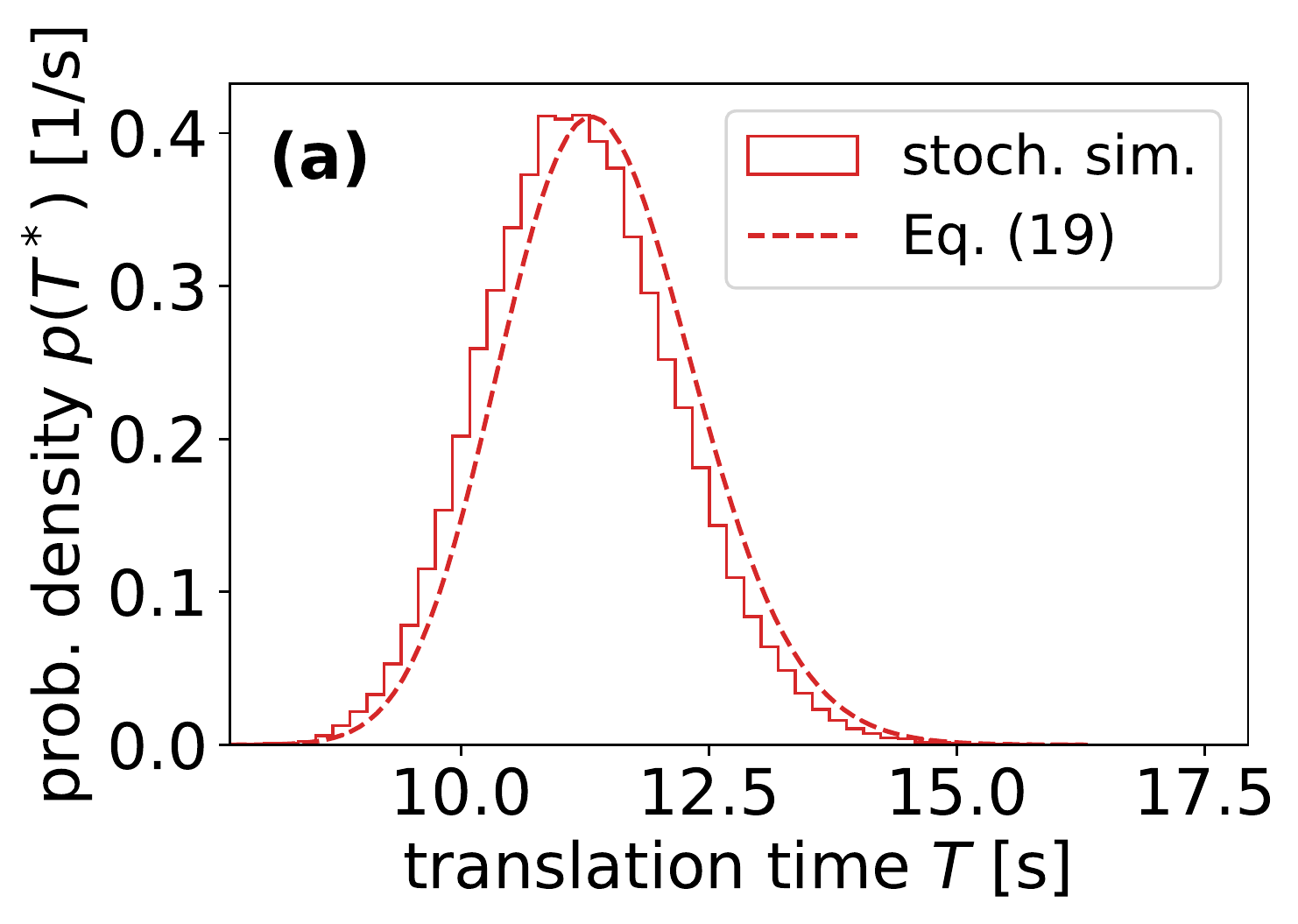}
	\includegraphics[height=1.55in]{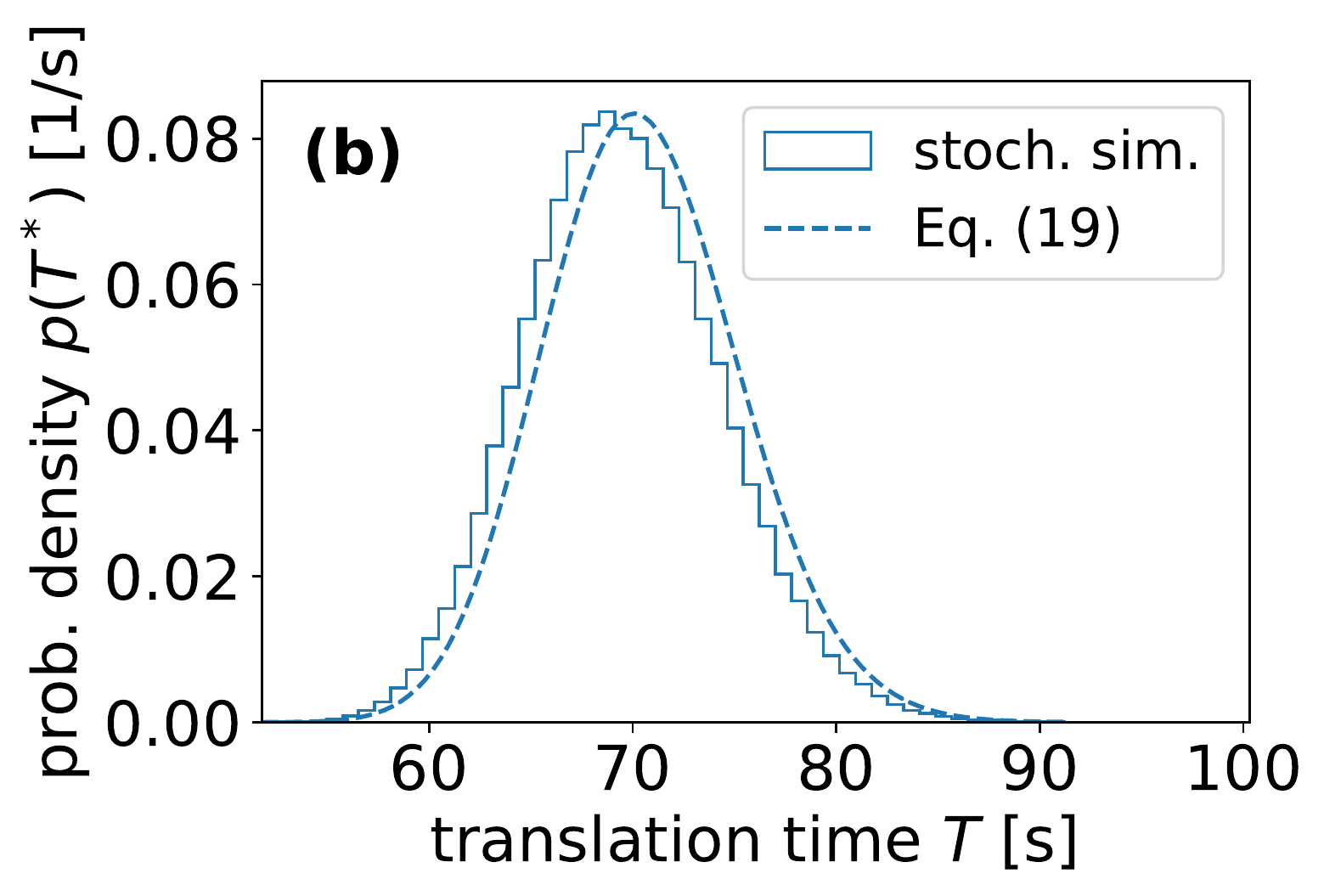}
	\includegraphics[height=1.55in]{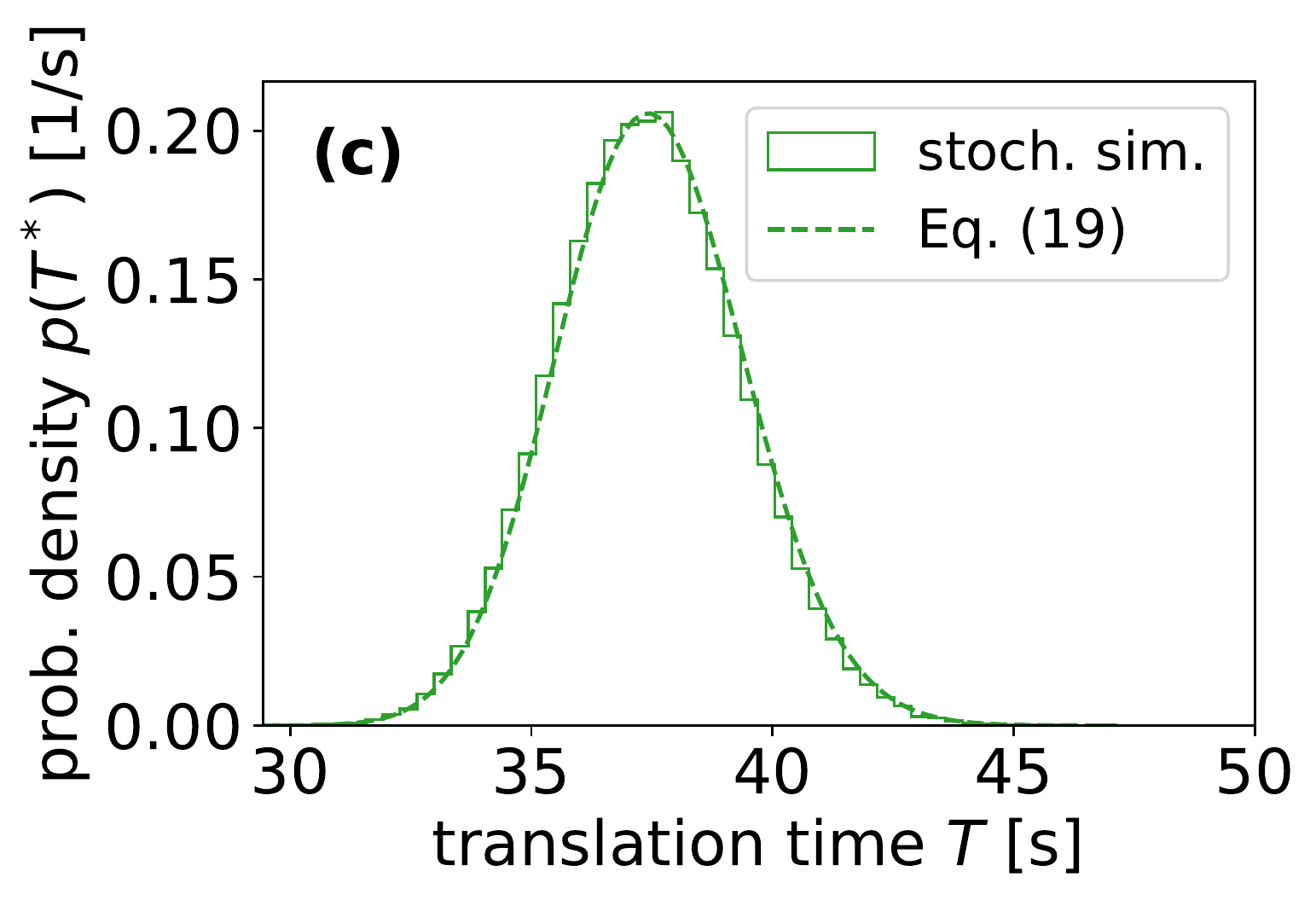}
	\caption{\label{fig7} Probability density function $p(T^*)$ of the translation time $T^*$ in the steady state for genes (a) sodA, (b) YAL020C and (c) beta-actin. The histograms are the result of stochastic simulations averaged over $10^5$ independent runs. Dashed lines were computed using Eq.~(\ref{pT}).}
\end{figure*}

\subsection{Distribution of the translation times \texorpdfstring{$T^*$}{T*}}

Within the effective medium approximation, the probability density function of the translation time $T^*$ in the steady state is equal to
\begin{equation}
\label{pT}
p(T^*)=\sum_{k=2}^{L}\lambda_k\textrm{e}^{-\lambda_k T^*}\left(\prod_{j=2, j\neq k}^{L}\frac{\lambda_j}{\lambda_j-\lambda_k}\right),
\end{equation}
while the mean and the variance of $T^{*}$ are given by
\begin{equation}
\label{t_s}
\langle T^*\rangle=\sum_{j=2}^{L}\frac{\rho_{j}^{*}}{J^*},\quad\langle T^{*2}\rangle-\langle T^*\rangle^2=\sum_{j=2}^{L}\left(\frac{\rho_{j}^{*}}{J^{*}}\right)^2.
\end{equation}
From here we can compute the average elongation rate in the steady state defined as 
\begin{equation}
v^*=\frac{(L-1)}{\langle T^*\rangle}\;,
\label{average-elongation-speed}
\end{equation}
which leads to a simple expression that depends only on $J^{*}$ and $\rho^*$,
\begin{equation}
v^{*}=\frac{J^*}{\rho^*}.
\label{average-elongation-speed2}
\end{equation}
We mention that the average translation time $\langle T^*\rangle$ and elongation speed $v^*$ have been recently computed by Sharma, Ahmed and O'Brien \cite{Sharma2018} using stochastic simulations for thousands of genes of \textit{E. coli}, \textit{S. cerevisiae} and \textit{H. sapiens}. 

Interestingly, we can combine Eqs.~(\ref{average-elongation-speed}) and (\ref{average-elongation-speed2}) into the following equation,
\begin{equation}
N^*=J^{*}\langle T^*\rangle\;,
\label{Little-law}
\end{equation}
which predicts that the long-term average number of ribosomes on the transcript ($N^{*}=\rho^{*}(L-1)$) is equal to the average rate at which ribosomes initiate translation ($J^{*}$) multiplied by the average time that a ribosome spends on the lattice ($\langle T^{*}\rangle$). In  queuing theory, this relationship is known as Little's law \cite{Little1961} and has been shown to be universal with respect to the details of the queuing process. A similar relationship has been used in fluid dynamics where the total amount of fluid in a given volume is equal to the residence time of a particle in that volume multiplied by the fluid influx; for a rigorous derivation of this law in stochastic lattice gases including the homogeneous TASEP see Ref. \cite{Zamparo2019}.

In Fig~\ref{fig7} we compare the probability density function $p(T^*)$ obtained by stochastic simulations to Eq.~(\ref{pT}) predicted by the effective medium approximation. The best agreement is found for beta-actin gene, while a small but visible disagreement is found for YAL020C and sodA genes. The excellent agreement for beta-actin is expected, because the average number of ribosomes per mRNA is only $1.2$ and therefore ribosome collisions are rare. However, the difference between $p(T^*)$ obtained by stochastic simulations and Eq.~(\ref{pT}) for YAL020C and sodA genes is puzzling. One might think that the difference is due to increased ribosome traffic caused by higher initiation rates relative to beta-actin gene, however the situation is more complex. For example, when the initiation rates of YAL020C and sodA genes are increased to $10$ s$^{-1}$,  the average translation time $\langle T^{*}\rangle$ is still  accurately described by Eq.~(\ref{t_s}), but the predicted distribution is much broader than the one from stochastic simulations. Interestingly, when we repeat the analysis for other genes (aaeA and ccmE) at the same high initiation rates, the difference between $p(T^*)$ obtained by stochastic simulations and Eq.~(\ref{pT}) becomes negligible. These findings suggest that the accuracy of the effective medium approximation at high initiation rates depends not only on the overall ribosome density and traffic but also on codon-specific elongation rates and their distribution along the mRNA sequence. 

\subsection{Dynamics of individual (tagged) ribosomes}

We now show that the effective medium approximation allows us to describe the kinetics of the tagged ribosome and find the distribution of its position $X(t)$.  A ribosome that is at codon position $X(t)=n$,  must have arrived arrived at $n$ at some earlier time $t'\leq t$ and have been waiting at $n$ for at least $t-t'$. The probability $P(X(t)=n\vert X(0)=2)$ that the ribosome is at position $X(t)=n$ at time $t$ given that it was at $X(0)=2$ at time $t=0$ can be then computed from
\begin{align}
& P(X(t)=n\vert X(0)=2)\nonumber\\
& \quad =\int_{0}^{t}dt'p(e_{2}^{*}+\dots+e_{n-1}^{*}=t')\int_{t-t'}^{\infty}dt''p_n(t'')\nonumber\\
& \quad = \int_{0}^{t}dt'p(e_{2}^{*}+\dots+e_{n-1}^{*}=t')\textrm{e}^{-\lambda_n(t-t')}.
\end{align}
The last expression is a convolution, in which case the Laplace transform of $P(X(t)=n\vert X(0)=2)$ is equal to the Laplace transform of $p(e_{2}^{*}+\dots+e_{n-1}^{*}=t')$ (see Appendix \ref{appendix_a}) multiplied by $1/(\lambda_n+s)$, the Laplace transform of $\textrm{exp}(-\lambda_nt'')$,
\begin{align}
& \int_{0}^{\infty}dt P(X(t)=n\vert X(0)=2)\textrm{e}^{-st}\nonumber\\
& \quad = \left(\prod_{j=2}^{n-1}\frac{\lambda_j}{\lambda_j+s}\right)\frac{1}{\lambda_n+s}=\frac{1}{\lambda_n}\prod_{j=2}^{n}\frac{\lambda_j}{\lambda_j+s}.
\end{align}
The product in the last expression the Laplace transform of $p(e_{2}^{*}+\dots+e_{n}^{*})$ (the sum now includes $e_{n}^{*}$), so that the final expression for the distribution of $X(t)$ is
\begin{align}
& P(X(t)=n\vert X(0)=2)\nonumber\\
& \quad =\frac{1}{\lambda_n}\sum_{i=2}^{n}\lambda_i\textrm{e}^{-\lambda_i t}\left(\prod_{j=2, j\neq i}^{n}\frac{\lambda_j}{\lambda_j-\lambda_i}\right).
\end{align}
This result can be generalised to any starting point $X(0)=m\leq n$, which will become handy in the next Section,
\begin{align}
\label{Pm}
& P(X(t)=n\vert X(0)=m)\nonumber\\
& \quad =\frac{1}{\lambda_n}\sum_{i=m}^{n}\lambda_i\textrm{e}^{-\lambda_i t}\left(\prod_{j=m, j\neq i}^{n}\frac{\lambda_j}{\lambda_j-\lambda_i}\right).
\end{align}
If all the effective rates are equal, $\lambda_i=\lambda$, the above expression is replaced by
\begin{equation}
\label{Poisson}
P(X(t)=n\vert X(0)=m)=\frac{(\lambda t)^{n-m}\textrm{e}^{-\lambda t}}{(n-m)!},
\end{equation}
which is the Poisson distribution.

\section{Run-off translation after inhibition of translation initiation}

We assume that the system is initially in the steady state, so that the probability to find a ribosome at codon $i$ is equal to $\rho_{i}^{*}$. At time $t=0$,  translation initiation is inhibited  (e.g. by harringtonine), which is equivalent to  setting the rate of initiation $\alpha$ to zero. Eventually the remaining ribosomes run off leaving an empty mRNA. Our goal is to find time evolution of $\rho_i(t)$ and $\rho(t)$ as they decrease to zero. 

Since translation initiation is inhibited after $t>0$, only ribosomes that are positioned at $j\leq i$ at $t=0$ will contribute to $\rho_i(t)$. If we treat each of these ribosomes as an individual tagged particle we can write the ribosome density at $i$ at $t$ as a sum over contributions from each of the tagged particles
\begin{equation}
\rho_i(t)=
\sum_{j=2}^{i}P(X(t)=i \vert X(0)=j)\rho_{j}^{*}.
\end{equation}
We now  approximate $P(X(t)=i \vert X(0)=j)$ using the effective medium approximation (\ref{Pm}), yielding a simple approximation for $\rho_i(t)$ and in turn the total density $\rho(t)$
\begin{align}
& \rho_i(t)=
\sum_{j=2}^{i}\frac{\rho_{j}^{*}}{\lambda_i}\sum_{k=j}^{i}\lambda_k\textrm{e}^{-\lambda_k t}\prod_{\substack{m=j\\m\neq k}}^{i}\frac{\lambda_m}{\lambda_m-\lambda_k},\label{rho_t_i_runoff}\\
& \rho(t)=\frac{1}{L-1}
\sum_{i=2}^{L}\sum_{j=2}^{i}\frac{\rho_{j}^{*}}{\lambda_i}\sum_{k=j}^{i}\lambda_k\textrm{e}^{-\lambda_k t}\prod_{\substack{m=j\\m\neq k}}^{i}\frac{\lambda_m}{\lambda_m-\lambda_k}.\label{rho_t_runoff}
\end{align}

\begin{figure}[hbt]
	\centering
	\includegraphics[width=3.4in]{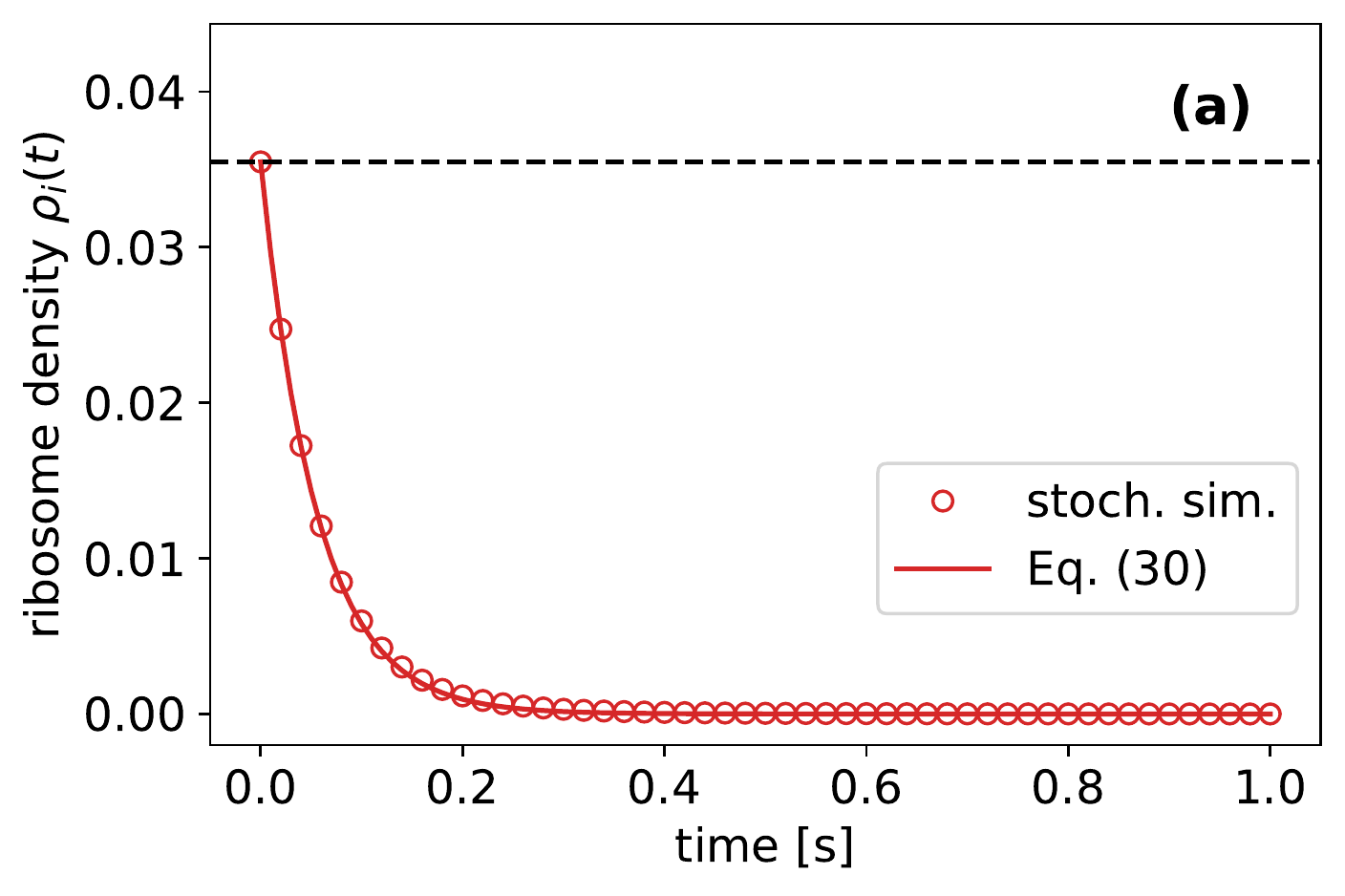}
	\includegraphics[width=3.4in]{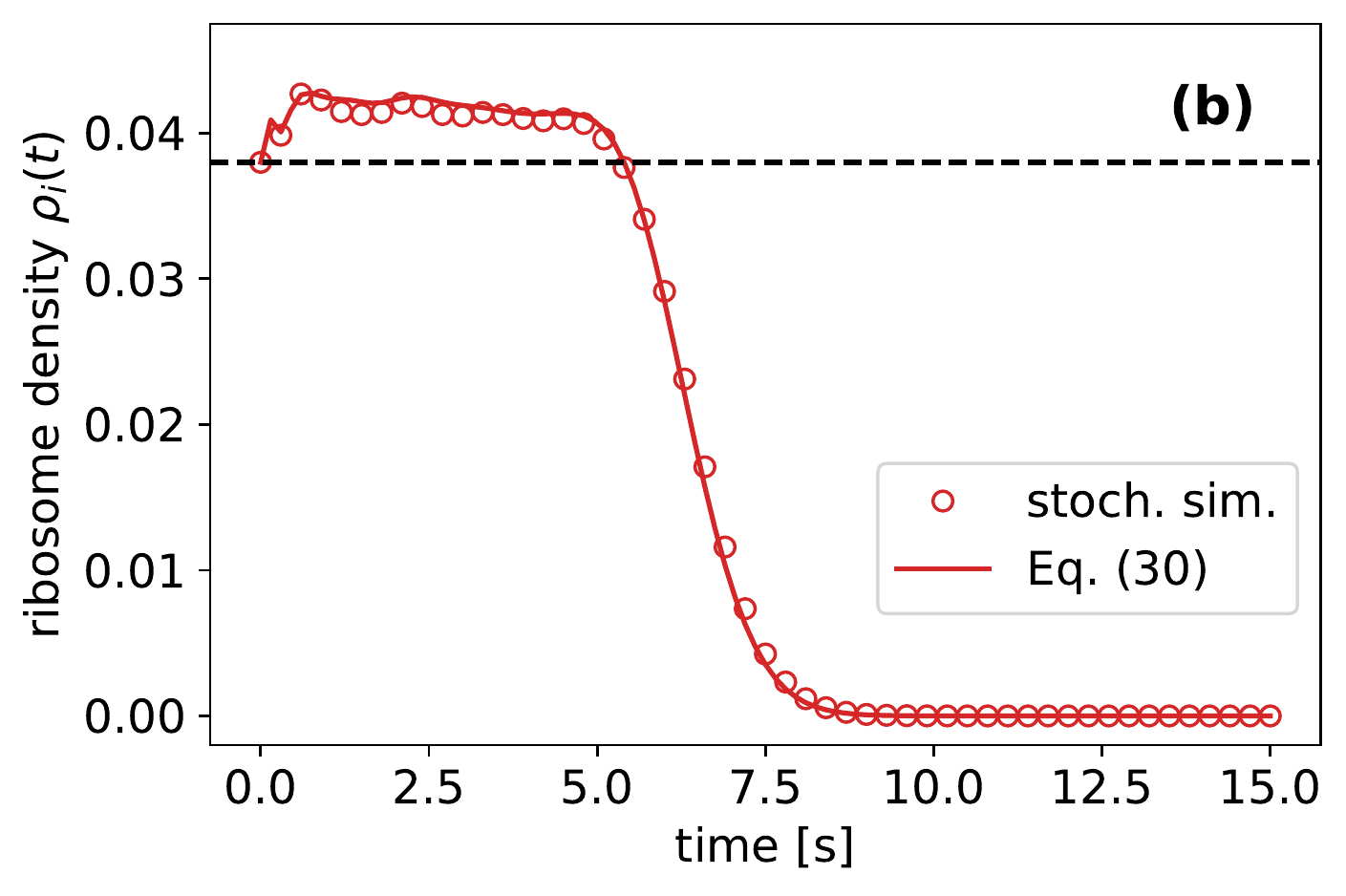}
	\caption{\label{fig8} Time evolution of the ribosome density $\rho_i(t)$ for gene sodA at codon positions (a) $i=2$ and (b) $i=100$ after translation initiation inhibition at $t=0$. Symbols are from stochastic simulations averaged over $10^6$ runs. Solid line was computed from Eq.~(\ref{rho_t_i_runoff}). Dashed horizontal line is the steady-state density $\rho_{i}^{*}$.}
\end{figure}

\begin{figure}[htp]
	\centering
	\includegraphics[width=3.4in]{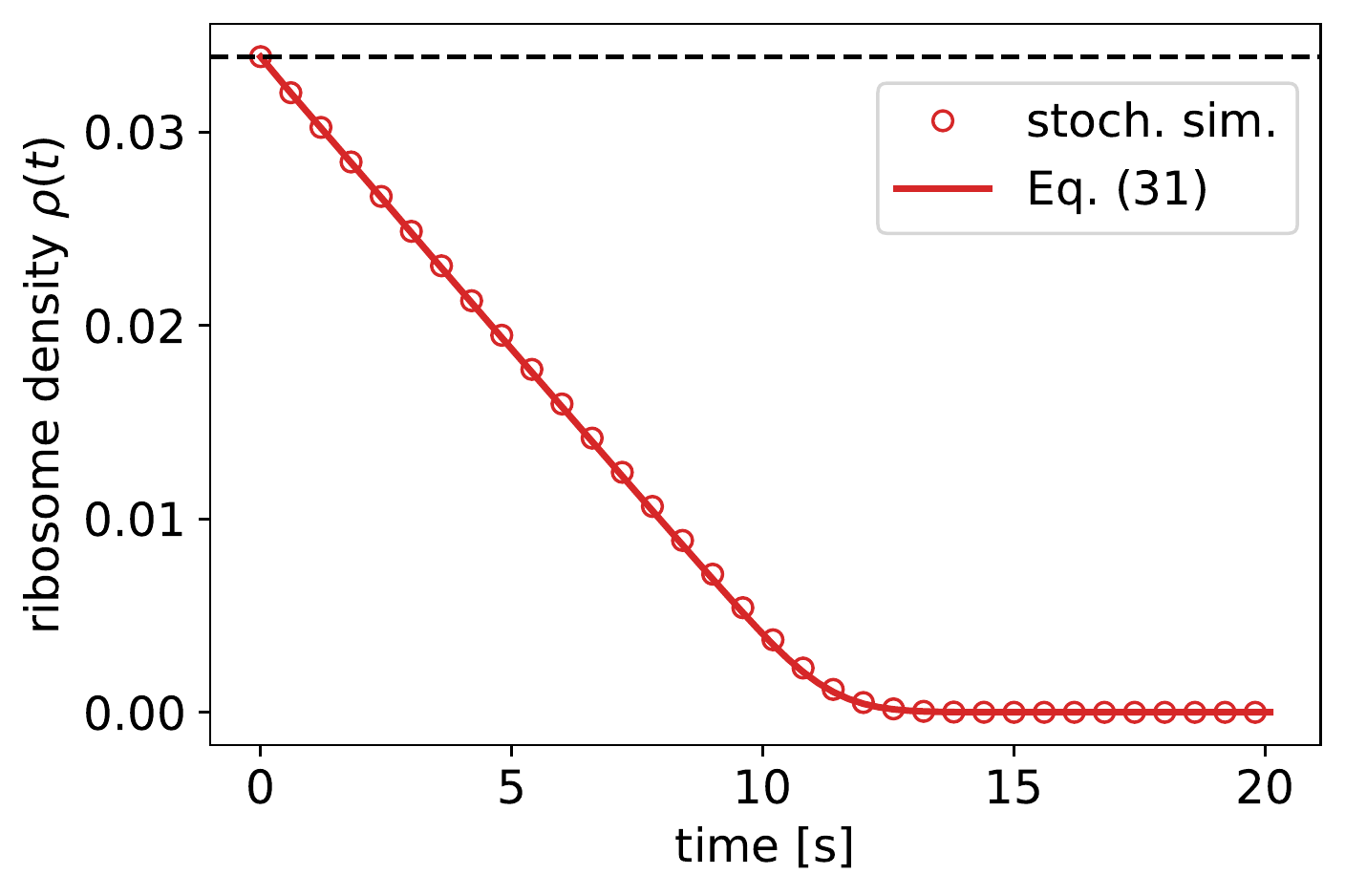}
	\caption{\label{fig9} Time evolution of the total ribosome density $\rho(t)$ after translation initiation inhibition at $t=0$ for gene sodA. Symbols are from stochastic simulations averaged over $10^{6}$ runs. Solid line was computed from Eq.~(\ref{rho_t_runoff}). Dashed horizontal line is the steady state density $\rho^{*}$.}
\end{figure}

To test the accuracy of this approximation we plot in Fig.~\ref{fig8} the time evolution of the ribosome density $\rho_i(t)$ obtained by stochastic simulations and compare to Eq.~(\ref{rho_t_i_runoff}) for sodA gene at two codon positions, $i=2$ and $i=100$. The time-dependent ribosome density $\rho_i(t)$ is accurately predicted by Eq.~(\ref{rho_t_i_runoff}) at both codon positions. For $i=100$, the early-time behaviour of $\rho_i(t)$ is non-monotonic due to non-uniform steady-state densities $\rho_{i}^{*}$ in Eq.~(\ref{rho_t_i_runoff}). Remarkably, the approximation captures this behaviour rather well. Finally, in Fig.~(\ref{fig9}) we demonstrate that the total ribosome density $\rho(t)$ predicted by Eq.~(\ref{rho_t_runoff}) reproduces the linear decay characteristic of run-off experiments \cite{Morisaki2016,Yan2016}. 

The total run-off time on a single mRNA is equal to the time it takes the last ribosome to leave the mRNA. If the last ribosome was at codon position $k$ at $t=0$, then the total run-off time is equal to $e_k+\dots+e_{L}$ and its average value is 
\begin{equation}
t_{\textrm{run-off}}(k)=\rho_{k}^{*}/J^*+\dots+\rho_{L}^{*}/J^*\quad \textrm{(single mRNA)}.
\end{equation}
In experiments, run-off traces are typically averaged over many mRNAs. In that case we need to weight $t_{\textrm{run-off}}(k)$ by the steady-state probability $P^{*}(\tau_2=\dots=\tau_{k-1}=0,\tau_k=1)$ that the last ribosome is at codon position $k$,
\begin{align}
t_{\textrm{run-off}}&=\sum_{k=2}^{L}P^{*}(\tau_2=\dots=\tau_{k-1}=0,\tau_k=1)\nonumber\\
&\quad\times t_{\textrm{run-off}}(k).
\end{align}
We can compute $P^{*}(\tau_2=\dots=\tau_{k-1}=0,\tau_k=1)$ using the power series method if translation is rate-limited by initiation \cite{Szavits2018,Scott2019} or using the mean-field approximation in the general case \cite{MacDonald1968,MacDonald1969,Shaw2003}.

We now consider the homogeneous TASEP \cite{Shaw2003}, where we can find closed form expressions for $\rho_i(t)$ within our effective medium approximation. First we summarise the steady state properties in the low-density phase:
\begin{align}
& \rho_{i}^{*}\approx\rho^{*}=\frac{\alpha}{\omega+\alpha(\ell-1)},\\
& J^{*}=\frac{\alpha(1-\alpha/\omega)}{\omega+(\ell-1)\alpha},\\
& \lambda_i=J^{*}/\rho_{i}^{*}=1-\alpha/\omega.
\end{align}
After inserting Eq.~(\ref{Poisson}) into Eq.~(\ref{rho_t_i_runoff}) the expression for $\rho_i(t)$ takes a simpler form
\begin{equation}
\label{rho_t_i_runoff_hom}
\rho_{i}(t)=\frac{\alpha}{\omega+\alpha(\ell-1)}\frac{\Gamma(i-1,(1-\alpha/\omega)t)}{\Gamma(i-1)},
\end{equation}
where $\Gamma(n)=(n-1)!$ and $\Gamma(n,x)$ is the upper incomplete Gamma function,
\begin{equation}
\Gamma(n,x)=\Gamma(n)\textrm{e}^{-x}\sum_{j=0}^{n-1}\frac{x^j}{j!}.
\end{equation}

\section{Conclusion}

In this paper we have modelled experiments that monitor translational kinetic within the framework of inhomogeneous TASEP. We have proposed simple and practical approximations which provide quantitative predictions consistent with observed phenomenology.\vspace{.3\baselineskip}

\noindent{\em Translation of a newly transcribed mRNA}. We have observed that the newly transcribed mRNA reaches the steady state shortly after the first, pioneering round of translation, provided translation is rate-limited by initiations i.e. $\alpha$ is small compared to the $\omega_i$. Thus the translation time for the pioneering ribosome may be taken as a proxy for the relaxation time into the steady state. We determine the full distribution of the first-round translation time and from there a simple expression for the time evolution of the ribosome density. In the context of mRNA degradation these findings imply that the steady state is reached {\rm before} the mRNA is degraded  (assuming that an mRNA has allowed the  production of at least one protein)---an assumption that is often made in the TASEP framework. In future work it would be of interest to determine whether the first-round translation time is correlated with the lifetime of mRNA molecule. \vspace{.3\baselineskip}

\noindent{\em Translation in the steady state}. In the steady state the dynamics of a tagged ribosome becomes more complicated due to collisions with other ribosomes. We circumvent this difficulty by introducing an effective medium approximation, which maps the tagged particle problem to a single particle problem with effective elongation rates. This approximation allows us to obtain the full distribution of the translation time in the steady state. In addition we find a simple expression for the average elongation rate
and the time-dependent probability for the tagged particle's position.\vspace{.3\baselineskip}

\noindent{\em Run-off translation after inhibition of initiation}. Initiation is switched off by setting $\alpha$ to zero, after which the remaining translating ribosomes run off.  We are able to describe this dynamical process in terms of steady-state quantities and the effective elongation rates of the effective medium approximation. Our results reproduce the time-dependence of the ribosome density, $\rho_i(t)$ at codon position $i$ as well the linear decay of the total ribosome density.

In this work we have compared our predictions with stochastic simulations of three particular genes. It would be of interest to further test the predictions genome-wide for different organisms. Finally, the approximations we have used, in particular the effective medimum approximation, may be of utility in more general TASEP-based models.

\begin{acknowledgments}
JSN was supported by the Leverhulme Trust Early Career Fellowship under grant number ECF-2016-768.
\end{acknowledgments}

\appendix
\section{Hypoexponential distribution}
\label{appendix_a}

If $x_1,\dots,x_n$ are independent random variables with probability density functions $p_i(x_i)=\lambda_i\textrm{exp}(-\lambda_i x_i)$, then their sum 
\begin{equation}
V = \sum_{i=1}^{n}x_i
\end{equation}
follows a hypoexponential distribution. We derive the corresponding probability density function $p(V)$ in two ways, one that uses Laplace transform and the other that uses mapping to a Markov process. The former is more appropriate for calculating moments and the latter is more suitable for evaluating $p(V)$.

By definition, 
\begin{align}
\label{pMdef}
 p(V)&=\int_{0}^{\infty}\textrm{d}x_1\dots\int_{0}^{\infty}\textrm{d}x_n\prod_{i=1}^{n}p_i(x_i)\nonumber\\
&\quad \times\delta\left(V-\sum_{j=1}^{n}x_j\right)
\end{align}
The Laplace transform $g(s)$ of $p(V)$ is equal to
\begin{equation}
\label{gs}
g(s)=\int_{0}^{\infty}\textrm{d}V\textrm{e}^{-sV}p(V)=\prod_{i=1}^{n}\frac{\lambda_i}{s+\lambda_i}.
\end{equation}
In order to find $p(V)$ we first perform the partial fraction decomposition of $g(s)$,
\begin{equation}
g(s)=\sum_{i=1}^{n}\frac{A_i}{s+\lambda_i}.
\end{equation}
Here we assumed that all $\lambda_i$ are distinct. The unknown coefficients $A_i$ can be found by multiplying $g(s)$ by $(s+\lambda_1)\dots(s+\lambda_n)$,
\begin{equation}
\sum_{i=1}^{n}A_i\prod_{j=1,j\neq i}^{n}(s+\lambda_j)=\prod_{j=1}^{n}\lambda_j.
\end{equation}
We now select $s=-\lambda_k$ so that only the term with $i=k$ survives in the sum yielding 
\begin{equation}
A_k=\lambda_k\prod_{j=1,j\neq k}^{n}\frac{\lambda_j}{\lambda_j-\lambda_k}.
\end{equation}
Finally, the probability density function $p(V)$ is given by
\begin{equation}
\label{pV-hypo}
p(V)=\sum_{i=1}^{n}\lambda_i\textrm{e}^{-\lambda_i V}\prod_{j=1,j\neq k}^{n}\frac{\lambda_j}{\lambda_j-\lambda_k}.
\end{equation}
From here the cumulative density function $P(V)$ is obtained by integrating $p(V)$ from $0$ to $V$,
\begin{equation}
P(V)=1-\sum_{i=1}^{n}\textrm{e}^{-\lambda_i V}\prod_{j=1,j\neq k}^{n}\frac{\lambda_j}{\lambda_j-\lambda_k},
\end{equation}
where we used that the sum of $A_i/\lambda_i$ over $i=1,\dots,n$ is equal to $g(0)=1$. In the special case in which all $\lambda_i=\lambda$, the resulting distribution is called the Erlang distribution and is probability density function reads
\begin{equation}
p(V)=\frac{\lambda^n V^{n-1}\textrm{e}^{-\lambda V}}{(n-1)!}.
\end{equation}

If we use Eq.~(\ref{pV-hypo}) to evaluate $p(V)$, we may run into problems with numerical precision. Namely, the product in Eq.~(\ref{pV-hypo}) can generate numbers smaller than the machine precision, which can lead to rounding errors. The solution is to write $p(V)$ using a matrix exponential, which can be computed using various algorithms.

The first step is to interpret $x_i$ as exponentially distributed waiting times in a Markov jump process in which  states $1,\dots,n$ are transient and state $n+1$ is absorbing. If $P_i(t)$ is the probability of being in state $i$ at time $t$, then 
\begin{equation}
\label{pM-Markov}
p(V)=\lambda_n P_n(V).
\end{equation}
The master equation for the probability $P_i(t)$ is given by
\begin{subequations}
\label{master-hypo}
\begin{align}
\frac{\textrm{d}P_1}{\textrm{d}t}&=-\lambda_1 P_1\\
\frac{\textrm{d}P_i}{\textrm{d}t}&=-\lambda_i P_i+\lambda_{i-1}P_{i-1},\quad i=2,\dots,n-1,\\
\frac{\textrm{d}P_n}{\textrm{d}t}&=-\lambda_n P_n + \lambda_{n-1} P_{n-1},
\end{align}
\end{subequations}
and the system is initially in state $1$, $P_i(0)=\delta_{i,1}$. We can write Eq.~(\ref{master-hypo}) as a first-order ordinary matrix differential equation,
\begin{equation}
\label{master-matrix}
\frac{\textrm{d}\mathbf{P}(t)}{\textrm{d}t}=\mathbf{M}\mathbf{P}(t),\quad\textbf{P}(0)\equiv\mathbf{a}=\begin{pmatrix}
1 \\
0 \\
\vdots\\
0
\end{pmatrix},
\end{equation}
where $\mathbf{P}$ is a column vector made of $P_i$, $\mathbf{M}$ is the following $n\times n$ matrix,
\begin{equation}
\mathbf{M}=\begin{pmatrix}
-\lambda_1 & 0 & 0 & \cdots & 0 & 0\\
\lambda_1 & -\lambda_2 & 0 & \cdots & 0 & 0\\
0 & \lambda_2 & -\lambda_3 & \ddots & 0 & 0\\
\vdots & \ddots & \ddots & \ddots & \ddots & \vdots\\
0 & 0 & \ddots & \lambda_{n-2} & -\lambda_{n-1} & 0\\
0 & 0 & \cdots & 0 & \lambda_{n-1} & -\lambda_{n}
\end{pmatrix},
\end{equation}
and $\mathbf{P}(0)$ is the initial probability vector which we denote by $\mathbf{a}$. The solution to Eq.~(\ref{master-matrix}) is
\begin{equation}
\mathbf{P}(t)=\textrm{e}^{t\textbf{M}}\textbf{a}.
\end{equation}
We can obtain $p(V)$ by adding Eqs.~(\ref{master-hypo}) together, which yields
\begin{equation}
\frac{\textrm{d}}{\textrm{d}t}\sum_{i=1}^{n}P_n=-\lambda_n P_n.
\end{equation}
Then we note that
\begin{equation}
p(V)=-\mathbf{1}^T\frac{\textrm{d}\mathbf{P}}{\textrm{d}t}=-\mathbf{1}^T\mathbf{M}\mathbf{P},
\end{equation}
where $\mathbf{1}$ is a column vector made of $1$ and $\mathbf{1}^T$ is the transpose of $\mathbf{1}$. The final expression for $p(V)$ is thus given by
\begin{align}
p(V)= -\mathbf{1}^T\mathbf{M}\textrm{e}^{V\mathbf{M}}\mathbf{a}.
\end{align}
If we now take a transpose of both sides we get the expression for $p(V)$ that is commonly found in the literature,
\begin{equation}
p(V)=-\mathbf{a}^{T}\textrm{e}^{V\mathbf{M}^T}\mathbf{M}^{T}\mathbf{1}.
\end{equation}
In order to find the cumulative distribution function $P(V)$ we use the identity
\begin{equation}
\left(\int_{0}^{V}\textrm{d}v\textrm{e}^{v\textbf{M}^T}\right)\textbf{M}^T=\textrm{e}^{V\textbf{M}^T}-\mathbf{I},
\end{equation}
where $\mathbf{I}$ is the identity matrix so that
\begin{equation}
P(V)=1-\mathbf{a}^{T}\textrm{e}^{V\textbf{M}^T}\textbf{1}.
\end{equation}
Thus both $p(V)$ and $P(V)$ can be computed at the same time by computing the matrix exponential $\textrm{exp}(V\mathbf{M}^{T})$.

\bibliography{references}

\end{document}